# Short-Range Order and Origin of the Low Thermal Conductivity in Compositionally Complex Rare-Earth Niobates and Tantalates

Andrew J. Wright [a], Qingyang Wang [b], Yi-Ting Yeh [a], Dawei Zhang [c], Michelle Everett [d], Joerg Neuefeind [d], Renkun Chen [b,c], Jian Luo [a,c,*]

[a] Department of NanoEngineering; [b] Department of Mechanical & Aerospace Engineering; [c] Program of Materials Science and Engineering, University of California San Diego, La Jolla, CA 92093, USA

[d] Neutron Sciences Directorate, Oak Ridge National Laboratory, One Bethel Valley Road, Oak Ridge, TN 37831, USA

## Abstract

Rare-earth niobates and tantalates possess low thermal conductivities, which can be further reduced in high-entropy compositions. Here, a large number of 40 compositions are synthesized to investigate the origin of low thermal conductivity. Amongst, 29 possess single (nominally cubic) fluorite phases and most of them are new compositionally complex (medium- or high-entropy) compositions. One new finding is that doping 2 % of light element cations can further reduce thermal conductivity. This large data set enables the discovery of a negative correlation between the thermal conductivity and averaged radius ratio of the 3+/5+ cations. While this ratio is still below the threshold for forming long-range ordered weberite phases, this correlation suggests the reduced thermal conductivity is related to short-range weberite order, which is indeed revealed by diffuse scattering in X-ray and neutron diffraction. Specifically, neutron diffraction is used characterize five selected specimens. A better fit to a weberite structure is found at nanoscale (~1 nm). The characteristic length (domain size) is smaller but with stronger short-range ordering in more insulative materials. As it approaches the Ioffe-Regel limit, the phonon limit breaks down and “diffusons” give rise to the observed amorphous-like thermal conductivity. Disordered oxygen sublattices are also confirmed by neutron diffraction.

*Correspondence should be addressed to J.L. (email: jluo@alum.mit.edu)

# Graphical Abstract

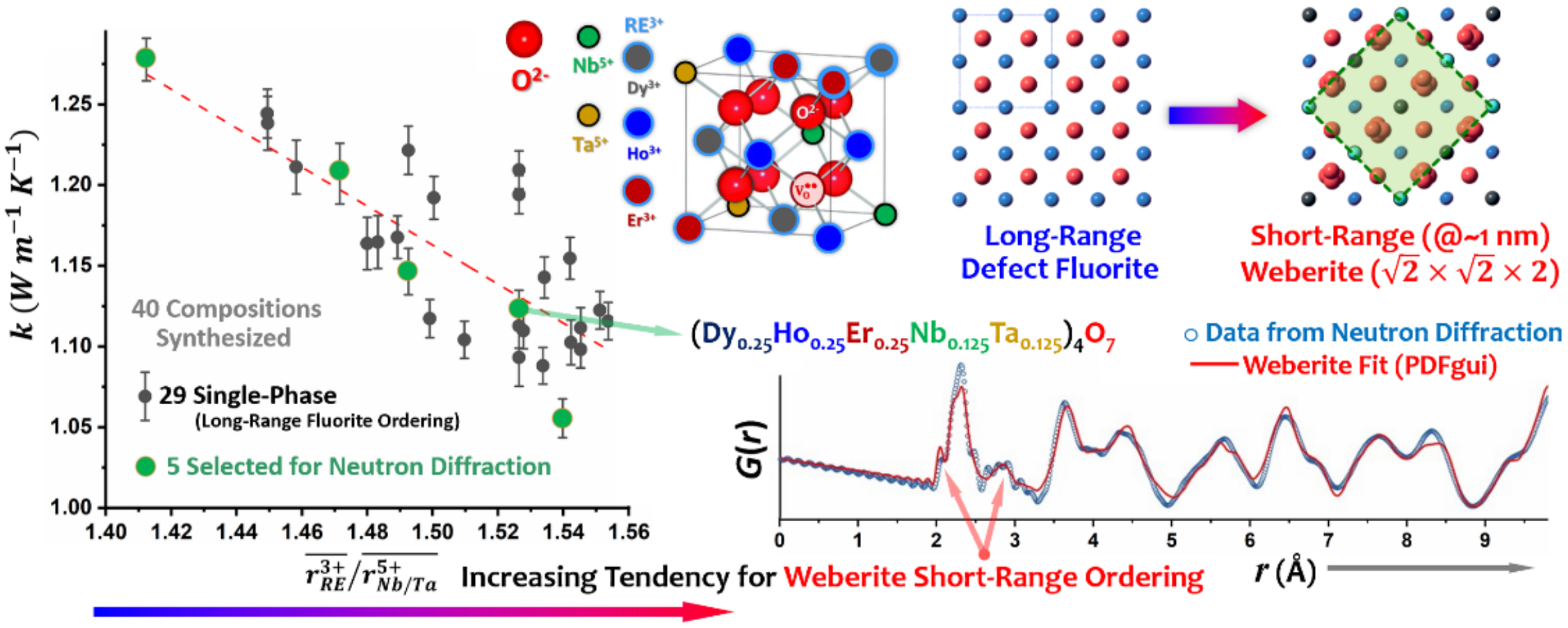

# 1 Introduction

Rare-earth (RE) niobates and tantalates ($RE_3NbO_7$ and $RE_3TaO_7$) are promising candidates for making the next generation thermal barrier coatings (TBCs) [1–6]. In general, developing new ceramic materials for TBCs is of critical importance for increasing the energy efficiency of gas-turbine engines [7,8]. Specifically, RE niobates and tantalates often form cubic "defect fluorite" structure (where $RE^{3+}$ and $Nb^{5+}/Ta^{5+}$ randomly occupy one cation sublattice, with a large fraction of 1/8th oxygen vacancies randomly distributed on the anion sublattice). An ordered weberite structure (of a $\sqrt{2} \times \sqrt{2} \times 2$ supercell with respect to the original cubic fluorite unit cell) can also form with a large radius ratio of the 3+ RE cation to the 5+ Nb or Ta cation ($r_{RE}^{3+}/r_{Nb/Ta}^{5+}$) [2,9]. These $RE_3NbO_7$ and $RE_3TaO_7$ possess high melting temperatures (typically >2000 °C), low thermal conductivities ($k$ < 2 $W\ m^{-1}K^{-1}$), and moderate Young's moduli ($E \sim 200 - 250\ GPa$) [1–6].

In the last a few years, the field of high-entropy ceramics (HECs) is developing rapidly and attracting great scientific interest [10–12]. The majority of recent studies of HECs have been focused on rock salt [13–15], perovskite [16–19], and fluorite [20–25], pyrochlore [26–28] structured oxides, metal diborides [29–32], and rock salt carbides [33–37]. Recent work also explored other systems such as other borides (including monoborides [38], $M_3B_4$ borides [39], tetraborides [40], and hexaborides [41]), silicides [42–44], aluminides [45], nitrides [46,47], fluorides [48,49], silicates [50–53], and phosphates [54,55]. Notable properties of interest include enhanced electrochemical [12,56–59] and mechanical [29–31,35,37,60–63] performances. Significant modeling efforts have also made to help understand the experimental results or guide the exploration of new compositions [35,63–66]. More recently, we further proposed to broaden HECs to compositionally complex ceramics (CCCs), where non-equimolar and/or medium entropy compositions can outperform their equimolar high-entropy counterparts [10,22,26]. Defects (*e.g.*, aliovalent cation doping or anion vacancies) and short- and long-range orders can bring additional complexity and opportunities to tailor the thermomechanical (and other) properties of these emerging HECs and CCCs [10].

Perhaps one of the most interesting and useful properties that HECs and CCCs possess, observed across nearly all systems examined, is represented by the reduced thermal conductivity [21–23,26,36,51,67–70]. Specifically, reduced and ultralow thermal conductivities have been shown in high-entropy RE niobates [10,23,71–73]. For example, in a 2020 perspective article [10], we reported the preliminary result for $(YDyErNb_{0.5}Ta_{0.5})O_7$ ($k$ = 1.11 $W\ m^{-1}K^{-1}$; $E$ = 226.5 GPa, NT1 in this study). Zhao *et al.* also reported the fabrication of high-entropy RE niobates and tantalates in 2020 (without reporting the thermal conductivity) [23]. A couple of recent articles in 2021 further reported ultralow thermal conductivities [71,72]. We also further investigated a series of single-phase duodenary oxides, *e.g.*, $[(Sm_{0.25}Eu_{0.25}Gd_{0.25}Yb_{0.25})_2(Ti_{0.5}Hf_{0.25}Zr_{0.25})_2O_7]_{1-x}[(Sc_{0.266}Dy_{0.248}Tm_{0.246}Yb_{0.240})_3NbO_7]_x$ with an order-disorder (pyrochlore-fluorite) transition, where one endmember is high-entropy RE niobates [73]. These recent studies collectively suggested that various high-entropy RE niobates and tantalates can be made into the (nominally cubic) fluorite structure (long-range order based on X-ray diffraction) with reduced and low thermal conductivities [10,23,71–73].

Braun et al. attributed low thermal conductivity in high entropy rock salt oxides to valency disorder inducing unit cell distortions [68]. Wright et al. showed the important roles of the concentration of oxygen vacancies and the size disorder in influencing the thermal conductivities in compositionally complex fluorite and pyrochlore oxides [22,26]. Here, we further investigate the underlying mechanisms of low thermal conductivities in high-entropy RE niobates and tantalates (including mixed niobates/tantalates).

In this study, we fabricated and investigated a large number of 40 RE niobates and tantalates (including 29 compositions that form single cubic phases), most of which are new compositionally complex (medium- and high-entropy) compositions not reported before. We examined the influence of the ratio of Nb and Ta and the effects of aliovalent doping. Furthermore, we introduced doping of light elements as a new route to further reduce thermal conductivity and tailor the thermomechanical properties. Notably, we discovered that the average $\overline{r_{RE}^{3+}}/\overline{r_{Nb/Ta}^{5+}}$ cation radii ratio controls the thermal conductivity. While the $\overline{r_{RE}^{3+}}/\overline{r_{Nb/Ta}^{5+}}$ ratios of our specimens are still well below the threshold for forming long-range ordered weberite phases, a negative correlation between the $\overline{r_{RE}^{3+}}/\overline{r_{Nb/Ta}^{5+}}$ ratio and thermal conductivity suggests the possible role of short-range weberite order, which was also evident by the diffuse scattering in X-ray diffraction. Furthermore, we employed advanced neutron diffraction, along with small-box modeling and large-box (reverse Monte Carlo) simulations, to characterize five selected specimens to reveal and confirm the existence of weberite nanodomains at the length scale of ~10 Å, as well as disordered oxygen sublattices. Thus, the observed amorphous-like ultralow thermal conductivity is likely to due to a "diffuson" mechanism [74].

## 2 Experimental Procedures

### *2.1 Materials Synthesis and Fabrication*

Constituent binary rare-earth oxide, $Al_2O_3$, and MgO powders (particle sizes ~5 μm) were purchased from US Research Nanomaterials. CaO (particle size < 160 nm), $Nb_2O_5$ (particle size ~ 500 nm), and $Ta_2O_5$ powders were purchased from Sigma-Aldrich, SkySpring Nanomaterials, and Inframat Advanced Materials, respectively. Stoichiometric amounts of binary oxides were weighed out with a 0.01 mg precision for a total of 2 g. The powders along with 2 wt% stearic acid was added to a poly(methyl methacrylate) tube with tungsten carbide endcaps and one Ø5/16" tungsten carbide ball. The powders were high-energy ball milled for 100 minutes (SPEX 8000D, SPEX SamplePrep, USA). The powders were then pressed under 100 MPa in a 0.5" diameter stainless steel die and placed on a Pt foil for sintering at 1600℃ or 1700℃ for 24 h. Both surfaces of the specimens were ground with a 30 μm diamond disc to remove any surface contamination. Pellets were ground into a powder with an agate mortar and pestle for diffraction experiments.

## *2.2 Characterization*

### *2.2.1 Phase, Density, and Compositional Uniformity*

X-ray diffraction (XRD, Miniflex II, Rigaku, Japan) was used to determine the crystal structure and theoretical density. Data was collected over 2 seconds per step with 0.02° 2θ steps. The bulk density was determined through the boiling method abiding the ASTM Standard C373-18 [75]. The relative density of the specimens ranged from 90 – 100 %.

Scanning electron microscopy (SEM, FEI Apreo, OR, USA) energy dispersive X-ray spectroscopy (EDS, Oxford N-Max$^N$) elemental mapping was used to characterize the compositional uniformity on cross-sectional specimens hot-mounted in acrylic (polished to 40 nm colloidal silica).

### *2.2.2 Neutron Diffraction*

Neutron diffraction experiments were carried out at Oak Ridge National Laboratory (ORNL) on the Nanoscale-Ordered MAterials Diffractometer (NOMAD, BL-1B) at the Spallation Neutron Source (SNS). Powders were loaded into a 2.8 $mm$ ID, 3 $mm$ OD diameter quartz capillary to a height of 3 $cm$ ($\sim 0.7\ g$). The diffraction experiments were carried out at $290\ K$ for a total accelerator proton charge of 8C, corresponding to about 92 min acquisition time at full power. The data from all six detector banks were background-subtracted (from an empty quartz capillary) and normalized to the intensity of vanadium prior to being combined to produce the total scattering function, $S(Q)$. This function was Fourier-transformed with a sliding $Q_{max}$ to obtain the neutron weighted pair distribution functions, $G(r) = r[g(r) - 1]$. The shortest distance correlations were transformed with a $Q_{max} = 45\text{Å}^{-1}$. The momentum transfer, $Q$, is given as $Q = 4\pi sin\theta/\lambda$, where $\theta$ and $\lambda$ are the scattering angle and neutron wavelength.

### *2.2.3 Young's Modulus (E)*

The Young's modulus was determined through an oscilloscope (TDS 420A, Tektronix, USA) operating in pulse-echo mode following the ASTM Standard C1198-20 [76]. The longitudinal and transverse wave speeds and thus, Poisson's ratio and Young's modulus were calculated by the same method used in a previous publication [22].

The measured Young's modulus was corrected for porosity according to $E = \frac{E_{measured}}{1-2.9P}$, where $P$ is the pore fraction [77]. All reported Young's moduli in this article are intrinsic materials properties extrapolated for fully dense specimens.

### *2.2.4 Thermal Conductivity (k)*

The thermal conductivity ($k$) was determined through the product of thermal diffusivity ($\alpha$), density ($\rho$), and heat capacity ($c_p$). Laser flash analysis (LFA 467 *HT HyperFlash*, NETZSCH, Germany) determined the thermal diffusivity. Before measurement, the specimens were coated with carbon to maximize laser absorption and infrared emission. The heat capacity was calculated through the weighted average of the heat capacity of the constituents [78] according to the Neumann-Kopp rule.

The measured thermal conductivity was corrected for porosity according $k = \frac{k_{measured}}{(1-P)^{3/2}}$, where $P$ is the pore fraction [79]. Again, all reported thermal conductivities in this article are intrinsic materials properties that represent the values extrapolated for fully dense specimens (with the porosity effects being corrected/removed).

## ***2.3 Neutron Diffraction Modeling***

### *2.3.1 PDFgui*

A small-box modeling approach was carried out with the PDFgui software [80]. A cif file was created in VESTA for each composition based on a $A_4O_7$ defect fluorite (space group 225, $Fm\overline{3}m$) and $A_3BO_7$ weberite (space group 20, $C222_1$) unit cell [81]. The $Q_{damp}$ and $Q_{broad}$ values were $0.017659$ Å$^{-1}$ and $0.0191822$ Å$^{-1}$, respectively. The fluorite cell had the lattice parameter, scale, correlated atomic motion (delta2), and the isotropic atomic displacement parameter (ADP) for the cation and anion site refined. The weberite cell had 27 parameters refined, including: lattice parameters, scale, correlated atomic motion (delta2), eight ADPs corresponding to the eight distinct lattice sites, and 14 lattice position parameters. The fitting range was selected to be from $0.02$ Å $- 70$ Å and the upper bound was then reduced in 5 Å steps until the final fitting range of $0.02$ Å $- 5$ Å. After each fitting window, the converged parameters were used as initial values for the next round of fitting.

### *2.3.2 Reverse Monte Carlo (RMC)*

A large-box modeling approach was carried out with the RMCprofile program [82]. Only the weberite cell was investigated using this technique because the weberite cell fit better in our PDFgui analysis (and previous researchers have also stated unsuccessful attempts in modeling $Yb_3TaO_7$ from a fluorite cell starting point [83]). A $6 \times 8 \times 8$ supercell was constructed for reverse Monte Carlo (RMC) analysis. Bank 2 from NOMAD was fitted in GSAS-II to provide an initial structure for the RMC simulation [84]. A Chebyshev background was manually determined so that the diffuse scattering is not captured. The lattice parameters, scale, atomic positions, ADPs, domain size, and microstrain were all refined.

Minimum bond lengths were selected for each atom-atom pair, but no maximum was set. The cations were allowed to move up to $0.05$ Å per translational move while O atoms were allowed up to $0.1$ Å. The rare earth cations all had an equal probability of swapping as did Nb and Ta (in compositions NT5-50 and NT5-50 4L). The probability for the simulation to generate translational or atomic swapping moves was approximately 50 %. The Bragg pattern from bank 2, $S(Q)$ ($F(Q)$ or $i(Q)$ in RMCprofile notation), and $G(r)$ ($D(r)_{normalized}$ in RMCprofile notation) were all fitted simultaneously. A simulation ran for $24\ h$ with no scaling, and then continued for another $24\ h$ with scaling. Post-analysis of the generated structure was carried out using CrystalMaker® 10.5 [85].

## 3 Results and Discussion

### *3.1 Phase Formation*

Among 40 compositions examined in this study, 29 exhibit a single (nominally cubic) fluorite phase and one exhibit single (orthorhombic) weberite phase according to the benchtop XRD analysis. Amongst, 26 single-phase solid solutions can be classified as medium- or high-entropy compositions (with 4-6 different metal cations and >1$k_B$ per cation in the ideal mixing configurational entropy [10]). The key results of all 40 compositions are documented in Table S1, along with all XRD spectra (Suppl. Figs. S4-S12), in the Supplementary Material (SM). The 29 single (nominally cubic) phase compositions and their room temperature thermal conductivity and Young's modulus are listed in Table 1. Selected SEM EDS elemental maps showing the compositional homogeneity are also documented in Suppl. Fig. S14-S15 in the SM.

Eight of the ten compositions that have secondary phases are those (non-conventional compositions) doped with 10% or more $TiO_2$, CaO, $AlO_{1.5}$ (= ½ $Al_2O_3$), and MgO (Suppl. Fig. S12a and S13 in the SM). Another is the NT10 doped with 2% $TiO_2$ with a trace amount of a secondary phase (Figure 2a). The only other two-phase composition is $Dy_3(Nb_{1/4}Ta_{3/4})O_7$, while $Dy_3TaO_7$ exhibits single orthorhombic weberite phase.

All the rest 29 conventional compositions with a variety of different RE combinations, including some doped with 2% CaO, $AlO_{1.5}$, and MgO, are single (nominally) cubic defect fluorite phase (for the long-range ordering based on XRD, albeit the possible existence of short-range weberite orders discussed later). Thus, this study suggests the ubiquitous formation of high-entropy RE niobates and tantalates.

### *3.2 Benchmark and Influence of Nb:Ta Ratio*

In a 2019 study, $Dy_3NbO_7$ was reported to have a ultralow thermal conductivity of $k = 1\ W\ m^{-1}\ K^{-1}$ and the highest reported *E*/*k* ratio [86]. An interesting experiment is to investigate the influence of substituting Nb with Ta (as Ta is approximately twice the mass of Nb, while Ta and Nb have almost the same ionic radius of ~0.64 Å [87]). Interestingly, $Dy_3NbO_7$ has a fluorite structure while $Dy_3TaO_7$ has a (long-range ordered) weberite structure; thus, a disorder-order transition, which can occur gradually with a two-phase transition region, can influence the properties. A recent study [88] of $Dy_3(Nb_{1-x}Ta_x)O_7$ ($x$ = 0, 1/6, 1/3, 1/2, 2/3, 5/6, and 1) found that the thermal conductivities are in the range of 1.1 $W\ m^{-1}\ K^{-1}$ for 100-900 °C, and both the modulus and thermal conductivity decrease with increasing Nb fraction. However, the fluorite-weberite transition complicates the interpretation.

Thus, we first investigated a similar series of compositions (NT0-0, NT0-25, NT0-75, and NT0-100 for $x$ = 0, 0.25, 0.5, 0.75, and 1 in $Dy_3(Nb_{1-x}Ta_x)O_7$) as a benchmark. Our $Dy_3NbO_7$ (*i.e.*, NT0-0, also expressed as $(Dy_{0.75}Nb_{0.25})_4O_7$ to emphasize that Dy and Nb randomly occupy the same cation sublattice in a defect $A_4O_7$ fluorite phase) has a measured thermal conductivity of $k = 1.10 \pm 0.03$ $W\ m^{-1}\ K^{-1}$ (10% higher than that reported in Ref. [86] and comparable with that reported in Ref. [88]) and modulus of $E$ = 201.8 ± 2.2 GPa. This yields a *E*/*k* ratio of 183.8 ± 2.8 $GPa\ m\ K\ W^{-1}$ (lower than that reported in Ref. [86]). The moderate differences may be attributed to the different measurements methods as well as the possible effects of residual porosity (while we note that our reported values represent intrinsic materials properties extrapolated for fully dense specimens). Overall, the agreements with literature are satisfactory, but we will focus on comparing the data obtained from our specimens fabricated and measured by the same procedure in most of the following discussion for consistence.

In our series of specimens, the thermal conductivity stays almost constant with increasing modulus for NT0-25 and NT0-50 with increasing Ta doping, producing higher $E/k$ ratios of 188-189 $GPa\ m\ K\ W^{-1}$ (Figure 1b). A (long-range) weberite phase started to form in $Dy_3(Nb_{1/4}Ta_{3/4})O_7$ (Figure 1a), where both modulus $E$ and conductivity $k$ increase abruptly (Figure 1b). The $E/k$ ratio dropped substantially in the weberite $Dy_3TaO_7$ due to a higher $k$ with slightly reduced $E$.

Next, we further investigated a series of compositionally complex $(Dy_{0.25}Er_{0.25}Ho_{0.25}Nb_{0.25-x}Ta_x)_4O_7$, which all show a single fluorite phase without long-range weberite ordering (Figure 1c). The measured $E$, $k$, and $E/k$ are shown in Figure 1(d). Both the modulus and conductivity increase as more Nb is substituted for Ta (albeit an initial drop of modulus with a small amount of Ta substitution), similar to the simpler $Dy_3(Nb_{1-x}Ta_x)O_7$ series. Notably, both $E$ and $k$ increase substantially from $x = 0.125$ to $x = 0.1875$ (corresponding to a change in an increase in Ta fraction) without a phase (long-range disorder-order) transition (Figure 1d). Interestingly, similar increases in $E$ and $k$ were observed in $Dy_3(Nb_{1-x}Ta_x)O_7$ accompanying a long-range ordering (fluorite-weberite phase transition) at the same Ta/(Nb+Ta) ratio (Figure 1b and Ref. [88]). Thus, these analogous abrupt increases in $E$ and $k$ leads one to wonder if short-range weberite ordering is emerged in compositionally complex $(Dy_{0.25}Er_{0.25}Ho_{0.25}Nb_{0.25-x}Ta_x)_4O_7$, despite the absence of the long range weberite order that occurs in the simpler composition of $Dy_3(Nb_{1-x}Ta_x)O_7$ with the Ta/(Nb+Ta) ratio greater than ¾.

### *3.3 Reducing Thermal Conductivity with Light Cation Doping*

As a new discovery, we further demonstrated that introducing a small amount of light element cations ($Mg^{2+}$, $Al^{3+}$, and $Ca^{2+}$) into a dense matrix of high-entropy tantalates, NT10 $(Er_{1/4}Tm_{1/4}Yb_{1/4}Ta_{1/4})_4O_7$, can further reduce the room-temperature thermal conductivity while maintaining single high-entropy solid-solution phases.

The 2% MgO, $AlO_{1.5}$ (1/2 $Al_2O_3$), and CaO doped $(Er_{1/4}Tm_{1/4}Yb_{1/4}Ta_{1/4})_4O_7$ were the single fluorite phase according to XRD, while a trace amount of secondary phase was observed in 2% $TiO_2$ doped $(Er_{1/4}Tm_{1/4}Yb_{1/4}Ta_{1/4})_4O_7$ (Figure 2a). In contrast, all the four 10% doped specimens had significant amounts of secondary phases (Suppl. Fig. S12a and Fig. S13 in the SM). The modulus of all the specimens decreases with doping (see Figure 2c for 2% doped compositions and all data in Suppl. Table S1 in the SM), which may be due to aliovalent doping and oxygen vacancy effects. The thermal conductivity decreases at 2% doping of MgO, $AlO_{1.5}$, CaO, and $TiO_2$ (Figure 2c). Most notably, 2% CaO doped $(Er_{1/4}Tm_{1/4}Yb_{1/4}Ta_{1/4})_4O_7$, which is single fluorite phase according to XRD (Figure 2a), possesses appreciably reduced thermal conductivity at all temperature range in comparison with $(Er_{1/4}Tm_{1/4}Yb_{1/4}Ta_{1/4})_4O_7$ (Figure 2c). In contrast, 2% doping of MgO, $AlO_{1.5}$, and $TiO_2$ reducing the room-temperature thermal conductivity, but increase the high-temperature thermal conductivity, of the $(Er_{1/4}Tm_{1/4}Yb_{1/4}Ta_{1/4})_4O_7$. The thermal conductivity increases significantly in the two-phase regime with 10% doping for all four light element oxide dopants (Suppl. Table S1 in the SM).

The reduction of the room-temperature thermal conductivity with 2% of light cations $Mg^{2+}$, $Al^{3+}$, and $Ca^{2+}$ in $(Er_{1/4}Tm_{1/4}Yb_{1/4}Ta_{1/4})_4O_7$ (without precipitating a secondary phases) represents an interesting new observation. This may suggest a "locon" effect (resulted from localized vibration modes of doped light element in a matrix of heavy elements) [74]. However, locons are usually the vibrational modes with the high frequency [74]. If it was the locon effect only, if would only impact the high frequency modes and would not impact the mechanical and acoustic properties (that are related to low-frequency modes). While the locon effect [74] remains a possible contributing mechanism, it is premature to draw a conclusion here. Instead, we hypothesize that the doping of the light element cations may have multiple effects. First, it does decease the Young's modulus, which will reduce phonon speed of sound (low frequency modes) and in turn decrease the thermal conductivity. Second, doping with light cations can increase the mass disorder (albeit that our correlation analysis shown in Suppl. Fig. S1b in the SM suggests the opposite: i.e., statistically, increased mass disorder surprisingly leads to slightly increased thermal conductivity in this class of materials). Third, some light element cations are also aliovalent, which can create charge defects and oxygen vacancies. In any case, the reduction of the thermal conductivity with small levels of light cations within the single high-entropy phases (Figure 2c) represents an interesting finding (and perhaps a new direction to further tailor the thermomechanical properties of high-entropy niobates and tantalates).

The decreases in the thermal conductivities in 10% doped specimens are likely related to the formation of significant amounts of secondary phases (as shown in Suppl. Fig. S12a and S13 in the SM) at the high doping level. These cases are too complicated and not sufficiently interesting to warrant further investigation.

### *3.4 Descriptor for Thermal Conductivity*

The successful fabrication of 29 single-fluorite-phase niobates and tantalates enable us to conduct data-driven analysis to identify the controlling factors and the best descriptor. We plotted the room-temperature thermal conductivity $k_{RT}$ versus five parameters ($\overline{r_{RE}^{3+}}/\overline{r_{Nb/Ta}^{5+}}$ ratio, size disorder $\delta_{\text{size}}$, mass disorder $g_{\text{mass}}$, density ρ, and ideal mixing entropy $\Delta S^{\text{mix,ideal}}$) in Figure 3 and Suppl. Fig. S1 in the SM and conducted correlation analyses to identify the controlling factors and the best descriptor.

Previously, we showed that a size disorder parameter ($\delta_{\text{size}} = \sqrt{\sum_{i=0}^{n} x_i \left(1 - {r_i}/{\bar{r}}\right)^2}$, where $x_i$ and $r_i$ are the mole fraction and ionic radius of the $i^{\text{th}}$ component and $\bar{r}$ is the weighted average ionic radius) controls the reduced thermal conductivity in compositionally complex pyrochlore oxides ($\delta_{\text{size}} = \sqrt{\left(\delta_{\text{size, A}}\right)^2 + \left(\delta_{\text{size, B}}\right)^2}$ for $A_2B_2O_7$ pyrochlores with two cation sublattices) [26]. In that case, a strong negative correlation between $\delta_{\text{size}}$ and $k_{RT}$ (*i.e.*, a larger $\delta_{\text{size}}$ implies a smaller $k_{RT}$) was also observed [26]. In the current case, a moderate negative correlation between $\delta_{\text{size}}$ and $k_{RT}$ with a Pearson correlation coefficient (PCC) of -0.61 was observed for these compositionally complex niobates and tantalates (Suppl. Fig. S1a in SM), which is consistent with the prior study [26]. It indicates that $\delta_{\text{size}}$ is still a (reasonably good) descriptor for forecasting reduced thermal conductivity, but it is less effective for compositionally complex niobates and tantalates than that for pyrochlore oxides [26].

Instead, we observed a stronger negative correlation between $\overline{r_{RE}^{3+}}/\overline{r_{Nb/Ta}^{5+}}$ and $k_{RT}$, as shown in Figure 3(a). Here, $\overline{r_{RE}^{3+}}$ is the weighted average effective ionic radius assuming a coordination number of VII according to Shannon [87] and $\overline{r_{Nb/Ta}^{5+}}$ is virtually a constant of 0.69 Å (since $Nb^{5+}$ and $Ta^{5+}$ have almost identical radii).

In addition, there is a rather complex relation between the mass disorder *vs.* $k_{RT}$ (Suppl. Fig. S1b in the SM) that is not easily interpretable. Statistically, increased mass disorder leads to slightly increased thermal conductivity, which is counterintuitive. There is a weak negative correlation between $k$ and density (PCC = -0.17, as shown in Suppl. Fig. S1c in SM).

Notably, the reduced thermal conductivity is not due to the "high entropy" effect; in contrast, we observed a weak positive correlation between $k$ and the ideal mixing entropy (PCC = 0.13, as shown in Suppl. Fig. S1d in SM). This correlation indicates an increasing mixing entropy in fact leads to increased $k_{RT}$ statistically (albeit that the correction is very weak and may not be statistically significant).

In summary, $\overline{r_{RE}^{3+}}/\overline{r_{Nb/Ta}^{5+}}$ appears to be most significant descriptor for forecasting $k_{RT}$ in compositionally complex niobates and tantalates based on the 29 single-phase compositions obtained in this study. Figure 2(a) shows that $k_{RT}$ reduces with increasing $\overline{r_{RE}^{3+}}/\overline{r_{Nb/Ta}^{5+}}$, or the increasing average size of the rare-earth cations $\overline{r_{RE}^{3+}}$ (since $\overline{r_{Nb/Ta}^{5+}}$ = 0.69 Å, virtually independent of the Nb/Ta ratio).

This interesting finding may be correlated with ordering in $RE_3NbO_7$ and $RE_3TaO_7$, where it is known that larger RE cations form an ordered weberite phase while the smaller cations (*e.g.*, Dy in $RE_3NbO_7$ and Ho in $RE_3TaO_7$) form the disordered fluorite phase [5,89]. Figure 3(b) expands Figure 3(a) to include compositions from literature [90–94]. It shows that long-range ordered weberite phases form at high $\overline{r_{RE}^{3+}}/\overline{r_{Nb/Ta}^{5+}}$ ratios (>~1.43), but they exhibit high thermal conductivities.

Our data generally agrees with previous reports for the fluorite oxides. However, the data from literature are more scattered. Some reports, e.g., Ref. [91] on a few ternary $RE_3NbO_7$ and a couple of high-entropy compounds, show surprisingly low thermal conductivities. These differences and scattered data are likely due to the differences in fabrications and measurements of thermal conductivities (*e.g.*, laser flash *vs.* hot-wire measurements), which can often have relatively large errors. In addition, the presence of residual porosity can often reduce the thermal conductivity appreciably. Thus, we mostly focus on comparing the relative values of our 29 single-phase compositions that were fabricated and measured via the same procedures, where we also carefully corrected the effects for porosity (so that all our data represent intrinsic thermal conductivities extrapolated for fully dense specimens).

The hypothesis that reducing thermal conductivity with increasing $\overline{r_{RE}^{3+}}/\overline{r_{Nb/Ta}^{5+}}$ is related to weberite order appears to be counterintuitive for the following two reasons. First, all the 29 single-phase compositionally complex niobates and tantalates are in the (at least nominally cubic) defect fluorite structure without long-range weberite ordering. Second, the long-range ordered (perfect) weberite phases typically have higher (instead of lower) thermal conductivities than defect fluorites. However, these apparently discrepancies can be fully explained and understood. We hypothesize that there are short-range weberite ordering in these compositionally complex niobates and tantalates, which maintain the long-range defect fluorite structure (nominally cubic based on XRD). Consequently, the formation of short-range orders, which have nanoscale characteristic size, can scatter phonons and reduce the thermal conductivity (in contrast to the increased thermal conductivity in the long-range ordered weberite phases).

To confirm and further analyze the existence of the short-range weberite ordering, we carried out XRD scattering and neutron diffraction on selective specimens. Five specimens labeled in Figure 3(a), with representative $k_{RT}$ and $\overline{r_{RE}^{3+}}/\overline{r_{Nb/Ta}^{5+}}$ values, were chosen for further in-depth study. These five selected compositions (labeled in Figure 3a) are:

- NT8: $(Sc_{0.25}Yb_{0.25}Lu_{0.25}Nb_{0.25})_4O_7$, $\overline{r_{RE}^{3+}}/\overline{r_{Nb/Ta}^{5+}}$ = 1.218; $k_{RT}$ = 1.28 ± 0.04 W m$^{-1}$ K$^{-1}$
- NT25: $(Sc_{0.165}Dy_{0.191}Ho_{0.197}Tm_{0.197}Nb_{0.25})_4O_7$, $\overline{r_{RE}^{3+}}/\overline{r_{Nb/Ta}^{5+}}$ = 1.337; $k_{RT}$ = 1.21 ± 0.03 W m$^{-1}$ K$^{-1}$
- NT4: $(Er_{0.25}Tm_{0.25}Yb_{0.25}Nb_{0.25})_4O_7$, $\overline{r_{RE}^{3+}}/\overline{r_{Nb/Ta}^{5+}}$ = 1.356; $k_{RT}$ = 1.15 ± 0.04 W m$^{-1}$ K$^{-1}$
- NT5-50: $(Dy_{0.25}Ho_{0.25}Er_{0.25}Nb_{0.125}Ta_{0.125})_4O_7$, $\overline{r_{RE}^{3+}}/\overline{r_{Nb/Ta}^{5+}}$ = 1.388; $k_{RT}$ = 1.12 ± 0.01 W m$^{-1}$ K$^{-1}$
- NT5-50 4L: $(Sc_{0.25}Yb_{0.25}Lu_{0.25}Nb_{0.25})_4O_7$, $\overline{r_{RE}^{3+}}/\overline{r_{Nb/Ta}^{5+}}$ = 1.218; $k_{RT}$ = 1.06 ± 0.01 W m$^{-1}$ K$^{-1}$

### *3.5 Temperature-Dependent Thermal Conductivity*

Figure 4 shows the thermal conductivities of five selected specimens up to 1000°C. Additional temperature-dependent thermal conductivities for all specimens are shown in Figure 2(b) for 2% light element cation doped compositions and documented in Suppl. Figs. S4(b)-S12(b) in the SM for all other compositions.

All five selected compositionally complex RE niobates and tantalates show amorphous-like conductivity trends, where $k$ increases slightly with temperature and plateaus at high temperatures (Figure 4) [74,95–97]. Such a behavior has been noted in simpler compositions such as $Dy_3NbO_7$ and appears to be common in fluorite-structured niobates and tantalates [92]. The minimum thermal conductivity ($k_{min}$) was computed for each specimen according to the model developed by Cahill, Watson, and Pohl [98]. The phonon limit ($k_{min}$) occurs when the mean free path of the phonon approaches half the phonon wavelength (Ioffe-Regel limit) and it is usually prevalent in highly disordered materials (that are chemically, structurally, or electronically disordered) [74,99,100]. Figure 4 shows that practically the thermal conductivities of all specimens lie below the $k_{min}$ at intermediate temperatures ($200 - 600$°C), while at high temperatures the thermal conductivity increases with temperature, possibly due to radiative contribution during laser flash measurements. Thermal conductivity lower than the $k_{min}$ indicates significant diffuson-mediated thermal transport as opposed to the typical phonon transport where the thermal conductivity variation with temperature shows $1/T$ trend [98]. Agne et al. derived an expression for the diffuson limit, which lies significantly lower than the $k_{min}$ derived by Cahill et al. [98]. Since the $k_{RT}$ values for the selected compositions are similar to the ternary compositions with identical charge distribution, we propose that the ultralow thermal conductivity in compositionally complex compositions is also largely driven by structural disorder.

### *3.6 Long-Range Order vs. Diffuse Scattering in XRD and Neutron Diffraction*

Benchtop XRD was carried out to determine the phase (long-range order). XRD patterns for five selected representative compositions are shown in Figure 5a) and additional XRD patterns for all compositions are documented in Suppl. Figs. S4(a)-S13(a) in the SM.

The Bragg peaks in Figure 5(a) correspond to a fluorite structure as expected. However, on closer inspection of the background, we noticed significant diffuse scattering occurring in all specimens around $2\theta$ = $40°$, $46°$, and 55°, which, in general, is the hallmark of correlated disorder [101–103]. We have shown this in Figure 5(b) on a logarithmic intensity scale for the five selected specimens with distinctly different conductivities and averaged RE cation radii. We also found that the magnitude of these broad peaks correlated with the measured $k_{RT}$ (or $\overline{r_{RE}^{3+}}/\overline{r_{Nb/Ta}^{5+}}$ ). Specifically, the smaller peaks were observed in compositions with lower conductivities and the more prevalent peaks were observed those with higher conductivities. This correlation suggests that a smaller peak corresponds to a smaller weberite domain (so the peak is broadened into the background noises) on the nanometer scale, which leads to a stronger scattering of phonons by the nanodomains resulting in a lower thermal conductivity.

These same five specimens were further characterized by neutron diffraction. Neutron diffraction is more sensitive to oxygen compared to X-rays and makes it a great complement in diffraction studies. The total scattering function $S(Q)$ is provided in Figure 5(c) and reveals more noticeable broad peaks in the background. A close-up of the background in Figure 5(d) reveals these same broad diffuse scattering peaks with higher clarity. The same trend discerned from XRD is also noticed here when examining the same peaks, namely, more thermally insulative specimens have smaller diffuse peak intensities.

Moreover, there are additional features in the neutron diffraction pattern that were not distinguishable in benchtop XRD. First, a large diffuse peak is found around 1 $Å^{-1}$ in $Q$-space in three of the compositions (NT25, NT5-50, and NT5-50 4L), but not in the other two (NT8 and NT4). A quick look at the first group of compositions reveals each of these materials contain the relatively large cations Dy and Ho. This peak was missed in XRD as it would have been near 15° in $2\theta$, which was below our scanning range. Second, the diffuse peak near 4.5 $Å^{-1}$ consists of two peaks in NT25, NT5-50, and NT5-50 4L (all containing Dy and Ho), whereas one broad peak exists in NT4 and NT8. This observation suggests that the materials with Dy and Ho have a stronger tendency for general ordering as seen by the increase in the number of peaks. In contrast, the materials without Dy and Ho (NT4 and NT8) may have a higher degree of ordering along certain orientations. In other words, the weberite domains in the materials containing Dy and Ho likely have spherical domains whereas the others likely have a stronger tendency to form planar domains. In all specimens, it was found that the frequently overlooked background contains rich information about diffuse scattering indicative of short-range order.

Similar diffuse scattering has been revealed by researchers in simpler rare-earth niobates and tantalates by Allpress and Rossell using electron diffraction patterns [104]. It was proposed then and confirmed later that the diffuse scattering corresponded to an orthorhombic weberite structure. Thus, the observed diffuse scattering in both XRD and neutron diffraction for compositionally complex niobates and tantalates suggests the existence of short-range weberite ordering (on the nanoscale) albeit that these materials show a long-range disordered defect fluorite structure.

### 3.7 Short-Range Ordering Indicated in Neutron Partial Distribution Function (PDF)

Figure 6 shows the PDF, $G(r)$, for all five specimens, which provides insight into the structure on the atomic scale in a histogram of the atom-atom distances. In general, each pattern had the same peaks and general shape alluding to the same general structure. On closer observation, it was found that the specimens containing Dy and Ho displayed a more ordered structure evident by the sharper peaks, especially around 3 Å. Specimen NT8, in particular, showed a very disordered structure with little distinguishable fine features. This agrees with the consensus that the larger the size difference between the 3+ and 5+ cations, the higher the tendency for ordering. This is similar to the relationship between the pyrochlore and fluorite structure in rare-earth zirconates and hafnates [105–107]. This trend is not followed for NT25 and NT4 since NT4 has a larger size ratio than NT25 (1.356 versus 1.337) yet has a more disordered PDF. This may suggest a tendency for short-range ordering among the 3+ cations.

We should note that Dy and Ho are strongly neutron absorbing elements, which may affect the accuracies of the PDFs of the samples that contain these elements.

### 3.8 Small-Box Modeling with PDFgui

The PDFgui software was used to investigate the local structure and a possible transition between the fluorite and weberite structures. We approached this by first fitting the PDF to a fluorite and weberite structure over a large radial distance (70 Å). This was to provide general lattice parameters that serve as an average over a relatively large number of unit cells ($\sim 13$ for the fluorite and $\sim 8$ for the weberite modeling). The maximum bound was decreased in 5 Å steps. This approached was followed to avoid any local minimums that may have resulted if this procedure was followed in reverse (i.e., increasing the maximum bound from 5 Å in 5 Å steps). The $R_w$ factor used to evaluate the fitting is given below:

$$R_w = \sum_{i=0}^{n} \left( \frac{w_i \left( Y_i^{obs} - Y_i^{calc} \right)^2}{\sum_{i=0}^{n} w_i \left( Y_i^{obs} \right)^2} \right)^{1/2} \times 100\% \qquad (1)$$

Here, $Y_i^{obs}$ and $Y_i^{calc}$ were the experimental and calculated data and $w_i$ is a weighting factor, which was set to unity. Additionally, the percent difference between the fluorite and weberite fitting factors were evaluated as a function of the fitting window.

Figure 7(a-e) shows the $R_w$ factor for the weberite and fluorite cells for each specimen. Across all materials, the weberite structure always produces lower $R_w$ at all observed fitting windows. This is likely due to two reasons. First, the weberite structure has been found to be a better fit for fluorite in the niobate and tantalate fluorite structure previously in short fitting windows [83]. Second, there are more degrees of freedom in the weberite structure so that it is easier to achieve smaller $R_w$. Nonetheless, they approach each other at sufficiently high radial distances. Thus, we focus on examining the trends in the change of $R_w$ values as functions of $r_{\max}$ (instead of the absolute values of $R_w$ for fitting fluorite vs. weberite structures), which provide more valid comparisons and information.

While the weberite structure always produces smaller $R_w$ than the fluorite structure for all cases (as discussed above), we focus on analyzing how the change of the relative value with changing $r_{\text{Max}}$. In the most disordered specimen, NT8 (Figure 7a), the difference (i.e., the preference to fit weberite structure over fluorite) increases only gradually with reducing $r_{\text{Max}}$. As the average size of the 3+ cations (or $\overline{r_{RE}^{3+}} / \overline{r_{Nb/Ta}^{5+}}$) increase, the weberite cell becomes more preferred structure (with a rapid decrease in $R_w$ for fitting the weberite structure, whereas an increase in $R_w$ for fitting the fluorite structure) at small radial distances ($r_{\text{Max}} = 5 - 10$ Å) and the difference between the two structures becomes smaller with increasing $r_{\text{Max}}$. This effect is particularly strong in NT5-50, where the weberite becomes the much more preferred fitted structure at $r_{\text{Max}} < 10$ Å. However, we note that this trend is slightly reversed (less significant) for NT5-50 4L (Figure 7e). While there is still a strong preference for the weberite structure at small radial distances ($5-10$ Å), it fit less well for NT5-50 4L than NT5-50. This may be due to the addition of La into the structure, making the rare-earth niobate/tantalate non-stoichiometric ($RE^{3+}$:(Nb, Ta)$^{5+}$ ≠ 3:1). This deviation from stoichiometry likely creates oxygen vacancies and promotes anti-site mixing, thereby resulting in more disordering than expected.

Overall, we can still conclude that weberite becomes the more preferred structure at the small length scale of $5-10$ Å, and this preference is more pronouncing in specimens with large average RE cation radii (or large $\overline{r_{RE}^{3+}}/\overline{r_{Nb/Ta}^{5+}}$) that is correlated to reduced thermal conductivity (Figure 7). In other words, this analysis suggests the occurrence of stronger short-range weberite ordering at the length scale of $\sim$10 Å in compositions with larger $\overline{r_{RE}^{3+}}/\overline{r_{Nb/Ta}^{5+}}$, which leads to lower thermal conductivity.

The percent difference between the fluorite and weberite fitting are shown in Figure 7(f-j). These plots further assist in showing the relative preference for weberite ordering in materials with larger rare-earth cations at small radial distances and a quicker approach to the fluorite fit. Again, we note that we focus on examining the trends in relative values as a function of $r_{\text{max}}$, but not the absolute values, for the rationale as discussed above). There is no strict cutoff (the relative values of fitted $R_w$) for the weberite-to-fluorite transition. If we take a somewhat arbitrary 20% cutoff in Figure 7(f-j), nanodomain sizes are likely only ~15-60 Å in most cases (Figure 7f-7j), and the nanodomain (or nanoscale ordering) becomes larger but less strong (more "diffuse") with reducing $\overline{r_{RE}^{3+}}/\overline{r_{Nb/Ta}^{5+}}$, particularly in NT8 (Figure 7f). Thus, they are not the traditional nanodomains (as the size is too small). Instead, the fluorite-structured $RE_3NbO_7$ and $RE_3TaO_7$ are the weberite structure at nearly the atomic scale, but the orthorhombic weberite cells are randomly distributed in three dimensions, which average out to a cubic cell at some characteristic length (corresponding to the proposed 20% cutoff), as proposed recently by researchers studying $Ho_2Zr_2O_7$ [9,105,106,108].

Using this definition (the proposed 20% cutoff), the "domain sizes" or characteristic lengths for the short-range weberite ordering for NT8, NT25, NT4, NT5-50, and NT5-50 4L are 60, 25, 50, 15, and 20 Å (as rough estimates; noting that the absolute values of these characteristic lengths may not be accurate because of the arbitrary selection of 20% cutoff, but their relative values and scale offer useful information). This concept still applies to thermal conduction as the vibrational density of states generated by this local weberite domain will be different than the outside bulk fluorite domain and attribute to phonon scattering. The domain size becomes increasingly small as the $\overline{r_{RE}^{3+}}$ increases. Notably, in NT5-50, the domain size approaches the distance of one weberite unit cell ($\sim 10.5 \times \sim 7.5 \times \sim 7.5$), which is where the definition of a phonon is no longer applicable [74,100]. At the Ioffe-Regel limit (phonon mean free path ~ $a$), the heat transfer is better described by a diffuson model where heat is carried through a random-walk model. This provides evidence as to why the thermal conductivity approaches and lies below the lower limit described by the phonon picture. It is noted that the trend holds up less well for NT25 and NT4. This again shows that it is unlikely to find one parameter to perfectly predict thermal conduction in these complex materials and other established phenomenon are still important. Here, we believe that the Sc present in NT25 provides a higher thermal conductivity than predicted by our domain size trend because Sc is a very light element and can transfer heat very efficiently due to high vibrational frequency it can achieve. In fact, all the compositions that contained Sc had the highest thermal conductivities among the niobates as seen in Table 1.

Moreover, we should note that the amorphous limit may be different for different compositions (e.g., higher in NT25). Thus, how the thermal conductivity is compared to the amorphous limit is more important than the absolute thermal conductivity, in term of understanding the underlying physics.

Additional effects such as anti-site defects and oxygen vacancies likely play a role in the reduced conductivity of NT5-50 4L, although it has a larger predicted domain size than NT5-50. An example of the fluorite and weberite fitting NT5-50 to a 10 Å limit using PDFgui is shown in Figure 8. The fluorite model is unable to capture peaks around 3 Å. The resulting $R_w$ was 29.0%. The weberite model improves this to 23.6% by capturing the general feature of nearest-neighbor region. PDFgui, in general, is a very simplified approach since it uses average parameters for a unit cell to model data. This has strong limitations in simple ternary niobates such as $Yb_3TaO_7$ [83]. This is expected to be exacerbated as the complexity of the composition increases, thereby requiring the use of more in-depth simulations.

### *3.9 Large-Box Modeling with RMCprofile*

The reverse Monte Carlo (RMC) method for fitting PDFs is highly revered as its ethos remains as a purely data-driven technique. The only restriction applied was that each atomic pair had a minimum distance. A $6 \times 8 \times 8$ weberite supercell (16896 atoms) was constructed with each cation randomly distributed among their sites. Another advantage this method has over small-box modeling is the capability to fit multiple datasets simultaneously, thereby being able to capture the structure on multiple length scales. Here, the PDF, $S(Q)$, and the diffraction from bank 2 were fit simultaneously. The fit using this method is shown in Suppl. Fig. S16 in the SM. The fit improves dramatically to a $R_w$ of 12.1%. However, we note that the RMC method can be prone to overfitting. Additionally, since the neutron data was transformed with a high $Q_{max}$ (45 Å$^{-1}$), which may result in a higher level of statistical fluctuations are present. Furthermore, Dy and Ho are relatively strong neutron absorbing elements, which may create unexpected anomalies within the PDFs. Thus, some of the intensity in the peaks around 3 Å may arise from statistical noises.

An examination into the nearest neighbor region along with the partial PDFs ($G_{ij}(r)$) for each specimen is shown in Figure 9. We found some difficulty in modeling the PDFs using this method due to the abundance of elements. This creates rather sharp peaks in the partial PDFs but produces a smoother total PDF that fits well to experiments. We had comparable results when we tried using the bond valence sum method as well, which is provided in the SM.

The first interactions in each composition are predominantly composed of Nb-O, Ta-O, and Sc-O while the RE-O interactions comprise of the large first peak. We found the peaks near the 3 Å region to correspond to O-O interactions. Thus, demonstrating that the larger cations induce more ordered oxygen sites such as in NT5-50 (Figure 9d) respective of an ordered weberite structure, whereas this is absent in NT8 (Figure 9a).

### *3.10 Atomic Density Variations*

The $6 \times 8 \times 8$ converged supercells were collapsed back down to one unit cell while retaining each atomic position to see the degree of disorder. A reference $Dy_3TaO_7$ from the ICSD database (collection code 191963) is provided in Figure 10(a), while each of the five selected high-entropy RE niobates or mixed niobates/tantalates is provided in Figure 10(b-d) in the $[001]$ viewing direction.

There are a few distinct observations here. First, NT25, NT5-50, and NT5-50 4L show numerous O atoms occupying the interstitials while this is minimal in NT8 and NT4. The location of these O atoms is random but tend to be localized to the region with the O-O nearest neighbor space. Second, specimens NT8 and NT4 show correlated disorder with the atomic positions. Both cation and (some) anion positions show stretching in the general direction of the (010) plane while this is not observable in the other specimens. It is also interesting to note that this stretching observed in the anion sublattice is asymmetrical along the $y$-axis. This feature likely arose from the absence of the low index diffuse scattering peak in Figure 5(d) at 1 $Å^{-1}$ that both NT8 and NT4 did not exhibit. The main weberite peak contributing to that diffuse scattering peak is the (110) plane, which is closely related to the stretching direction. Considering all cases, it is clear that there is extreme disorder in the O sublattice while the cations are relatively ordered. This further supports the evidence that the diffuse scattering peaks are enhanced by neutron diffraction in contrast to X-rays.

A heatmap of the atomic density was calculated for each model (including the reference) for the atoms in the $[001]$ plane at $z = 0.25$ in Figure 10(g-l). The transition from zero to maximum intensity is denoted by dark blue to white. Again, in specimens NT8 and NT4 (Figure 10h & 10j), a couple of the O sites are displaced along a direction similar to the $(010)$ but not exactly. The sites appeared to be more ordered in the other specimens. While NT25, NT5-50, and NT5-50 4L have less displacement within the sites, they instead have population of O within interstitials and can be distinguish in some of the heatmaps, particularly NT25 (Figure 10i). An alternative view of Figure 10 where the zone axis is slightly tilted in the $+y$ direction to illustrate depth is provided in Suppl. Fig. S17 in the SM.

The weberite structure is identical to the fluorite structure when $a/c = \sqrt{2}$ and $c = b$. The lattice parameters along with these two ratios are provided next to the heatmaps in Figure 10. All materials are very close to the cubic conditions, but the small deviations are what give rise to the diffuse scattering and weberite short-range ordering. While we do believe this is inherent to the material system (not necessarily a high-entropy feature) [83], we would like to note that it is unlikely for any high-entropy system to be perfect cubic system (free of short-range ordering). There should always remain a tendency for short-range ordering and deviations from the perfect cubic symmetry as each atom has slightly different affinities and size [105,109].

### *3.11 Elemental Site Distribution and Clustering*

This point was investigated further by considering the elemental site occupancy and possibility of clustering within the supercell. This was accomplished by taking three unique orientations and determining the elemental distribution of each column (accounting for all the atoms in the plane). The results for NT4 are shown in Figure 11 while the rest are in the Suppl. Figs. S18-S21 in the SM.

Specimen NT4 $(Er_{0.25}Tm_{0.25}Yb_{0.25}Nb_{0.25})_4O_7$ reveals a fairly constant distribution of Er across all axes within the supercell (Figure 11). However, it also shows poor site mixing of Tm and Yb. The A-site of the weberite structure alternates between CN-VII and CN-VIII sites. In this specimen, the large-box modeling approach suggests that Tm is preferred to be on the CN-VIII site while Yb is predominantly on the CN-VII site. This may be explained by Yb being able to change its electronic configuration to fill the $f$-orbital with more ease than Tm. In fact, all other specimens investigated in this study also show site preference on the A-site, although to lesser extents. In NT25, NT5-50, and NT5-50 4L, Dy and Ho have poor mixing with Dy chiefly on the VII site and Ho on the VIII site. In NT8, Sc and Yb show near equal distributions throughout the supercell, while Sc and Yb (mainly VII site) have poor mixing with Lu (mainly VIII site). Nb and Ta in all specimens show a random distribution within the B-site.

Our key finding here is that the presence of larger cations such as Dy and Ho drastically affect the structure of the O sublattice. The smaller cations composed within NT8 and NT4 demonstrate a structure resembling ordered sites, but highly anisotropic displacements (stretched) near the (010) direction. The specimens containing larger cations such as Dy and Ho (NT25, NT5-50, and NT5-50 4L) induce more of isotropically displaced O sublattice but has some O atoms occupying interstitials. It is yet unknown how this affects the thermal properties. While the structure of the O sublattice seems to be highly dependent on large cations such as Dy and Ho, it does not correlate well the measured thermal conductivity. However, the radius ratio, $\overline{r_{RE}^{3+}}/\overline{r_{Nb/Ta}^{5+}}$, appears to be a suitable descriptor that possibly accounts for all of these effects in a simple approximation. This is similar to how the radius ratio describing the fluorite-to-pyrochlore transition, $r_A^{3+}/r_B^{4+}$, embodies the complex phenomenon of anti-site mixing energetics [107]. Inelastic neutron scattering experiments are proposed to obtain the phonon dispersion curves to see how the observed correlated disorder versus random disorder affects phonon broadening.

## 4 Conclusions

Overall, 40 RE niobates and tantalates, including 29 single-phase defect fluorite compositions, were synthesized to advance the understanding of this emerging class of low thermal conductivity materials. Based on this rather large set of data, we found that the average size of the 3+ RE cations (or $\overline{r_{RE}^{3+}}/\overline{r_{Nb/Ta}^{5+}}$ cation radii ratio) is a good descriptor to forecast trends in room-temperature thermal conductivity.

Five selected representative compositions were chosen for further in-depth characterizations. The temperature-dependent thermal conductivity confirms the glass-like trend with value approaching or slightly falling below the phonon limit. XRD revealed a long-range order fluorite structure but indicated diffuse scattering peaks representative of the orthorhombic weberite structure in the background. Neutron diffraction increased the clarity of these diffuse scattering peaks. The neutron PDFs showed the lowest thermal conductivity specimens (which had the largest cations) were exhibit more short-range weberite ordering. A small-box modeling approach on these PDFs reveal the lower thermal conductivity materials fit much stronger to a weberite structure in the nearest-neighbor region compared to the fluorite structure, but it was quickly comparable to the fluorite cell once the fitting window was increased. This finding is consistent with a small weberite domain size approaching the Ioffe-Regel limit (*i.e.*, phonon limit). A reverse Monte Carlo approach was also tested and fit the PDFs better while also producing a supercell structure. Additionally, the large-box modeling efforts suggest certain elements (Dy, Yb, and Sc) to have preference to the VII site. While the structure is correlated on the presence of certain elements, it does not correlate well with the thermal conductivity. Instead, it is suggested that a simple descriptor, $\overline{r_{RE}^{3+}}/\overline{r_{Nb/Ta}^{5+}}$ , may embody various factors into a simple parameter to forecast trends in reduced thermal conductivity in this class of materials.

We also revealed that doping of small amount of light element oxides can further reduce the room-temperature thermal conductivity of single-phase high-entropy tantalates (and potentially also niobates), which suggests another route to further tailor the thermomechanical properties of compositionally complex RE niobates and tantalates via incorporating “nonconventional” dopants.

**Acknowledgment:** The project (on the fundamental research of the role of disorder and short-range ordering in influencing thermomechanical properties, including the neutron diffraction analysis of short-range ordering and its role in reduced thermal conductivity) is supported by the National Science Foundation (NSF) via Grant No. DMR-2026193. We also acknowledge a seed project (Small Innovative Projects in Solar) supported by the U.S. DOE Solar Energy Technologies Office (Contract No. EE0008529) where the materials fabrication and thermal conductivity measurements started before August 2020. A portion of this research used resources at the Spallation Neutron Source, a DOE Office of Science User Facility operated by the Oak Ridge National Laboratory.

**Supplementary Materials**

Supplementary material related to this article (including Suppl. Table S1 summarizing the key results of all 40 compositions investigated and Suppl. Figs. S1-S21) can be found, in the online version, at doi: xxxxxxxx

**Table 1.** The thermomechanical properties of each niobates and tantalates synthesized in this study sorted by $\overline{r^{3+}}/\overline{r^{5+}}$. Specimens examined by neutron diffraction are denoted by *.

| Identifier | Composition | $\overline{r^{3+}}/\overline{r^{5+}}$ | $k_{RT}\left(\frac{W}{m\cdot K}\right)$ | $E_{RT}\ (GPa)$ | $E/k\ (GPa\ m\ K\ W^{-1})$ |
|---|---|---|---|---|---|
| NT8* | $(Sc_{0.25}Yb_{0.25}Lu_{0.25}Nb_{0.25})_4O_7$ | 1.281 | 1.28 ± 0.04 | 225.2 ± 2.3 | 176.3 ± 2.6 |
| NT20 | $(Sc_{0.168}Er_{0.234}Tm_{0.183}Yb_{0.165}Nb_{0.25})_4O_7$ | 1.316 | 1.24 ± 0.02 | 211.4 ± 1.9 | 170.7 ± 2.8 |
| NT21 | $(Sc_{0.2}Dy_{0.186}Tm_{0.185}Yb_{0.18}Nb_{0.25})_4O_7$ | 1.316 | 1.24 ± 0.02 | 214.1 ± 1.9 | 172.1 ± 2.6 |
| NT24 | $(Sc_{0.185}Dy_{0.21}Er_{0.182}Yb_{0.173}Nb_{0.25})_4O_7$ | 1.324 | 1.21 ± 0.05 | 217.9 ± 1.8 | 179.9 ± 2.9 |
| NT25* | $(Sc_{0.165}Dy_{0.191}Ho_{0.197}Tm_{0.197}Nb_{0.25})_4O_7$ | 1.337 | 1.21 ± 0.03 | 217.8 ± 1.7 | 180.4 ± 3.2 |
| NT10-2A | $(Er_{0.245}Tm_{0.245}Yb_{0.245}Al_{0.02}Ta_{0.245})_4O_{7-\delta}$ | 1.343 | 1.18 ± 0.04 | 209.0 ± 2.7 | 179.6 ± 3.4 |
| NT10-2M | $(Er_{0.245}Tm_{0.245}Yb_{0.245}Mg_{0.02}Ta_{0.245})_4O_{7-\delta}$ | 1.347 | 1.20 ± 0.04 | 207.4 ± 2.7 | 177.6 ± 3.0 |
| NT26 | $(Sc_{0.15}Dy_{0.279}Ho_{0.171}Er_{0.15}Nb_{0.25})_4O_7$ | 1.347 | 1.16 ± 0.03 | 222.9 ± 1.9 | 191.4 ± 3.2 |
| NT4* | $(Er_{0.25}Tm_{0.25}Yb_{0.25}Nb_{0.25})_4O_7$ | 1.356 | 1.15 ± 0.04 | 217.9 ± 2.5 | 190.1 ± 3.2 |
| NT10 | $(Er_{0.25}Tm_{0.25}Yb_{0.25}Ta_{0.25})_4O_7$ | 1.356 | 1.22 ± 0.04 | 226.5 ± 3.1 | 185.4 ± 3.4 |
| NT6 | $(Y_{0.25}Tm_{0.25}Yb_{0.25}Nb_{0.25})_4O_7$ | 1.363 | 1.19 ± 0.04 | 215.6 ± 2.5 | 180.8 ± 2.9 |
| NT10-2C | $(Er_{0.245}Tm_{0.245}Yb_{0.245}Ca_{0.02}Ta_{0.245})_4O_{7-\delta}$ | 1.365 | 1.15 ± 0.03 | 204.1 ± 2.3 | 182.7 ± 2.8 |
| NT2 | $(Dy_{0.25}Er_{0.25}Yb_{0.25}Nb_{0.125}Ta_{0.125})_4O_7$ | 1.372 | 1.10 ± 0.03 | 223.9 ± 2.7 | 202.8 ± 3.3 |
| NT5 | $(Dy_{0.25}Ho_{0.25}Er_{0.25}Nb_{0.25})_4O_7$ | 1.388 | 1.09 ± 0.03 | 214.2 ± 2.4 | 196.0 ± 3.9 |
| NT5-25 | $(Dy_{0.25}Ho_{0.25}Er_{0.25}Nb_{0.188}Ta_{0.063})_4O_7$ | 1.388 | 1.11 ± 0.01 | 202.9 ± 2.2 | 182.4 ± 3.0 |
| NT5-100 | $(Dy_{0.25}Ho_{0.25}Er_{0.25}Ta_{0.25})_4O_7$ | 1.388 | 1.21 ± 0.01 | 222.4 ± 3.5 | 183.9 ± 3.4 |
| NT5-75 | $(Dy_{0.25}Ho_{0.25}Er_{0.25}Nb_{0.063}Ta_{0.188})_4O_7$ | 1.388 | 1.19 ± 0.01 | 224.2 ± 3.1 | 187.8 ± 3.2 |
| NT5-50* | $(Dy_{0.25}Ho_{0.25}Er_{0.25}Nb_{0.125}Ta_{0.125})_4O_7$ | 1.388 | 1.12 ± 0.01 | 207.6 ± 2.7 | 184.8 ± 3.0 |
| NT1 | $(Y_{0.25}Dy_{0.25}Er_{0.25}Nb_{0.125}Ta_{0.125})_4O_7$ | 1.389 | 1.11 ± 0.03 | 226.5 ± 2.6 | 204.1 ± 3.1 |
| NT5-50 2L | $(Dy_{0.245}Ho_{0.245}Er_{0.245}La_{0.02}Nb_{0.123}Ta_{0.123})_4O_{7-\delta}$ | 1.393 | 1.09 ± 0.01 | 222.7 ± 2.7 | 204.6 ± 3.3 |
| NT3 | $(Y_{0.25}Dy_{0.25}Ho_{0.25}Nb_{0.25})_4O_7$ | 1.395 | 1.14 ± 0.03 | 208.8 ± 2.2 | 182.7 ± 2.8 |
| NT5-50 4L* | $(Dy_{0.24}Ho_{0.24}Er_{0.24}La_{0.04}Nb_{0.12}Ta_{0.12})_4O_{7-\delta}$ | 1.399 | 1.06 ± 0.01 | 220.6 ± 2.5 | 209.0 ± 3.4 |
| NT32 | $(Y_{0.143}Dy_{0.587}Ho_{0.02}Nb_{0.143}Ta_{0.107})_4O_7$ | 1.403 | 1.15 ± 0.06 | 226.7 ± 3.0 | 196.4 ± 3.4 |
| NT5-50 6L | $(Dy_{0.235}Ho_{0.235}Er_{0.235}La_{0.06}Nb_{0.118}Ta_{0.118})_4O_{7-\delta}$ | 1.404 | 1.10 ± 0.01 | 218.9 ± 3.0 | 198.6 ± 3.7 |
| NT0-0 | $(Dy_{0.75}Nb_{0.25})_4O_7$ | 1.406 | 1.10 ± 0.03 | 201.8 ± 2.2 | 183.8 ± 2.8 |
| NT0-25 | $(Dy_{0.75}Nb_{0.188}Ta_{0.063})_4O_7$ | 1.406 | 1.11 ± 0.01 | 209.4 ± 2.4 | 188.4 ± 3.0 |
| NT0-50 | $(Dy_{0.75}Nb_{0.125}Ta_{0.125})_4O_7$ | 1.406 | 1.10 ± 0.01 | 207.6 ± 2.3 | 189.1 ± 2.9 |
| NT5-50 8L | $(Dy_{0.23}Ho_{0.23}Er_{0.23}La_{0.08}Nb_{0.115}Ta_{0.115})_4O_{7-\delta}$ | 1.409 | 1.12 ± 0.01 | 201.3 ± 2.2 | 179.3 ± 2.7 |
| NT5-50 10L | $(Dy_{0.225}Ho_{0.225}Er_{0.225}La_{0.1}Nb_{0.113}Ta_{0.113})_4O_{7-\delta}$ | 1.415 | 1.12 ± 0.01 | 208.8 ± 2.6 | 187.1 ± 3.0 |

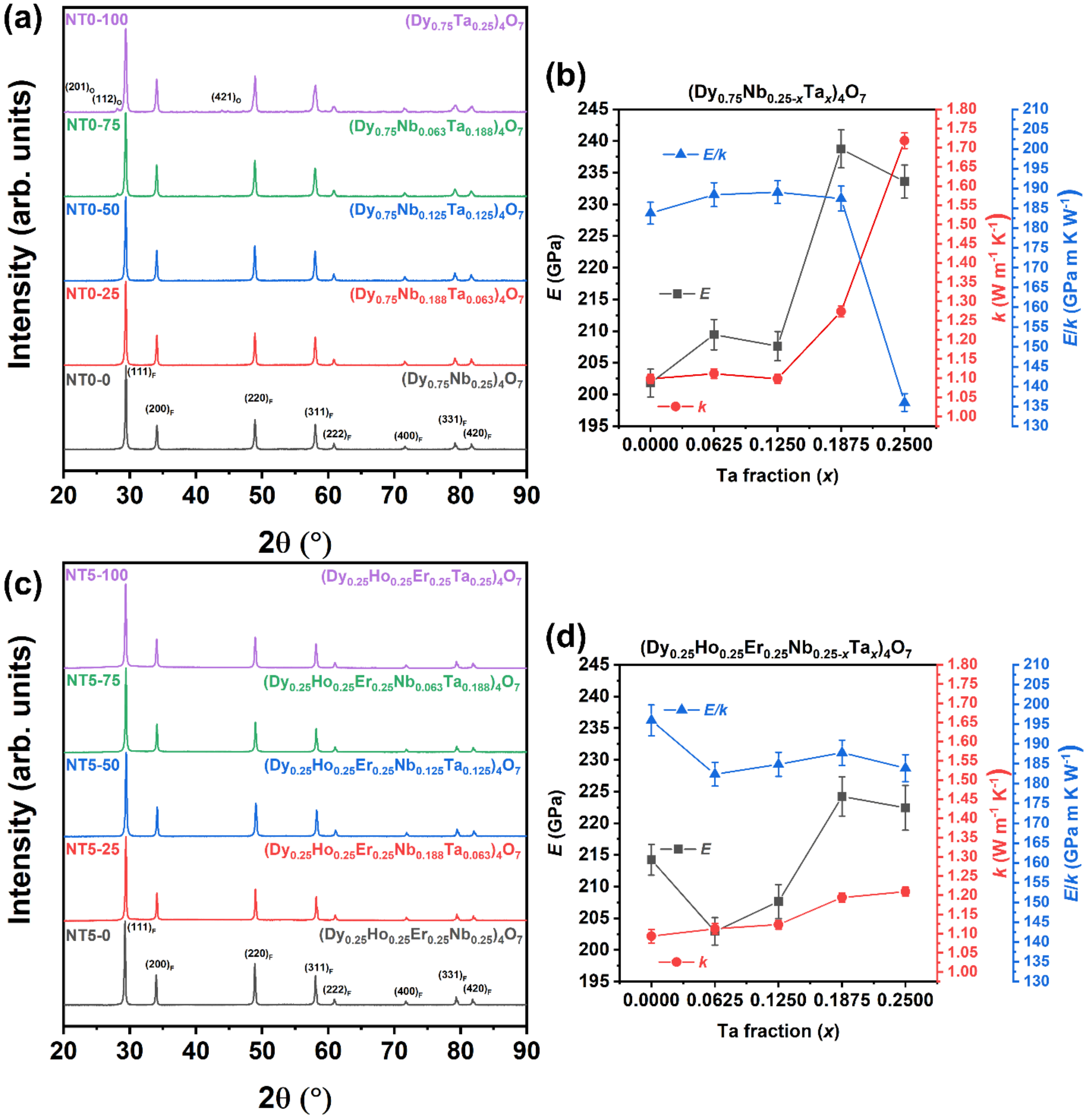

**Figure 1. (a)** XRD pattern evolution from fluorite $Dy_3NbO_7$ to orthorhombic $Dy_3TaO_7$. **(b)** The Young's modulus ($E$), thermal conductivity ($k$), and $E/k$ ratio for $Dy_3Nb_{1-x}Ta_xO7$ series. **(c)** XRD patterns and **(d)** the measured Young's modulus ($E$), thermal conductivity ($k$), and $E/k$ ratio of a $(Dy_{0.25}Er_{0.25}Ho_{0.25}Nb_{0.25-x}Ta_x)_4O_7$ series.

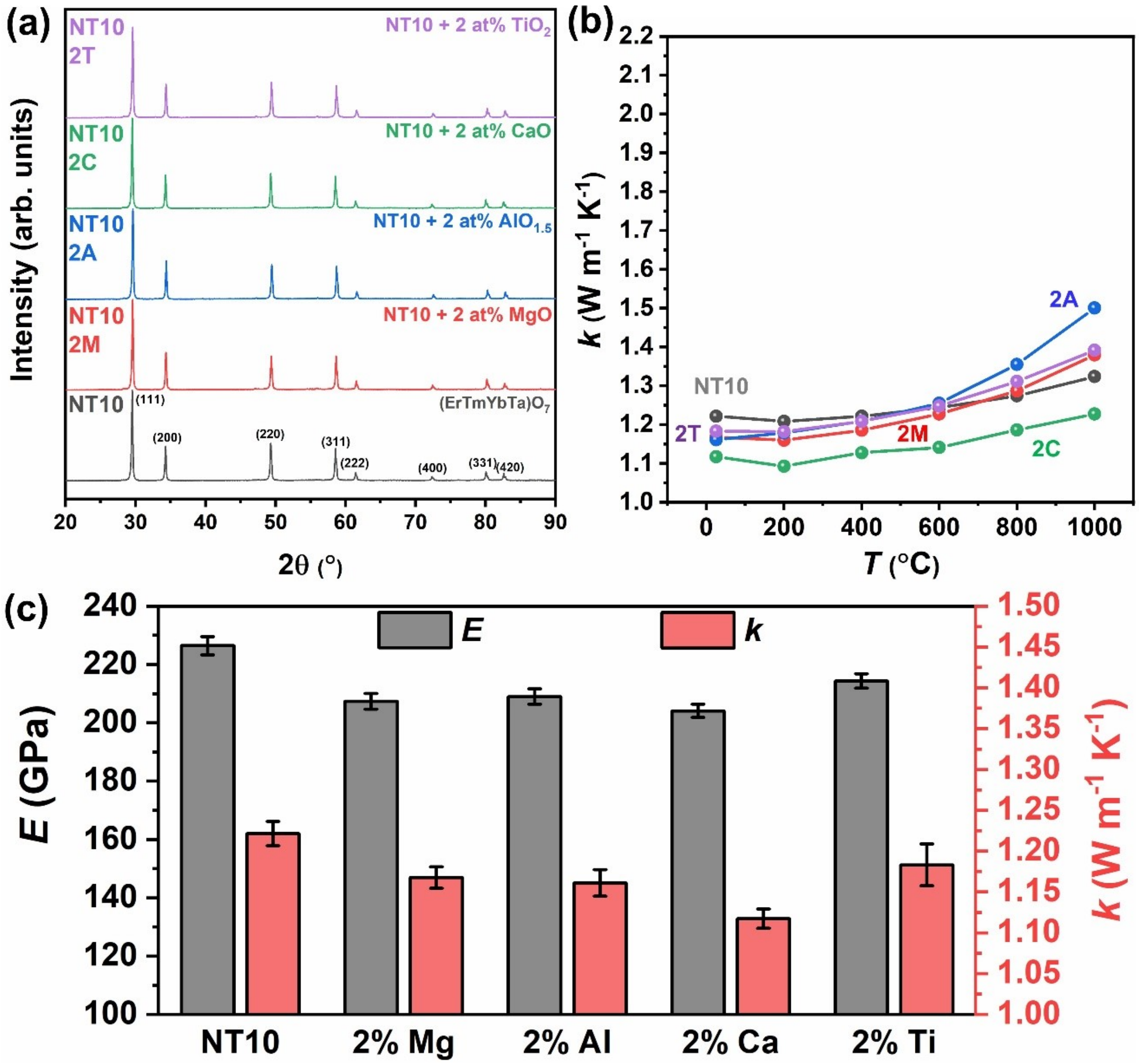

**Figure 2.** **(a)** Benchtop XRD spectra and **(b)** temperature-dependent thermal conductivity of NT10, $(Er_{1/3}Tm_{1/3}Yb_{1/3})_3TaO_7$, doped with 2 mol.% MgO, $AlO_{1.5}$, CaO, and $TiO_2$. **(c)** Room temperature Young's modulus and thermal conductivity for each specimen.

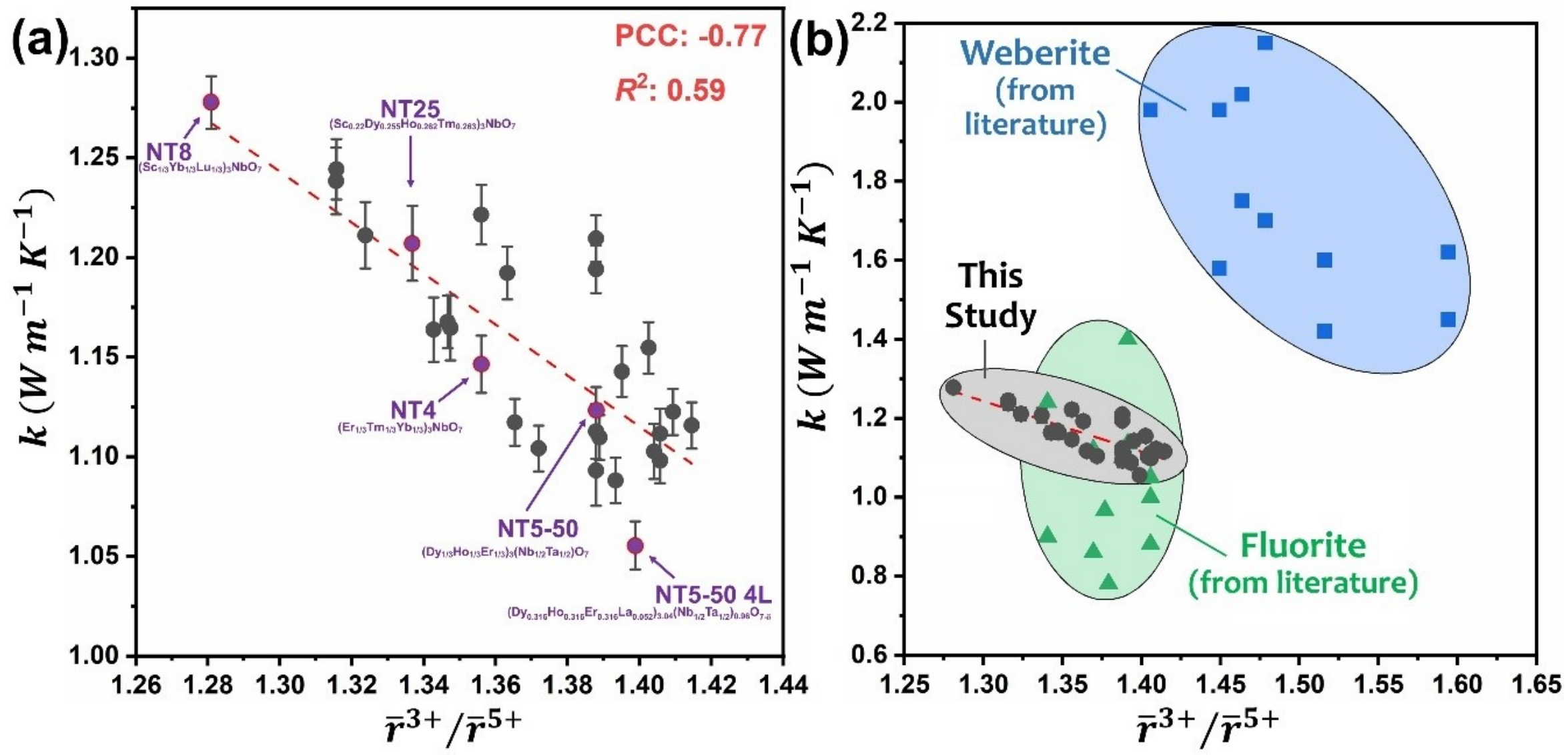

**Figure 3. (a)** Correlation of room temperature thermal conductivity *vs.* the radius ratio of the 3+ cations to the 5+ cations. The highlighted specimens were selected for the neutron diffraction study. **(b)** Data in panel (a) expanded to include data of (long-range) fluorite and weberite structures from Refs. [90–94].

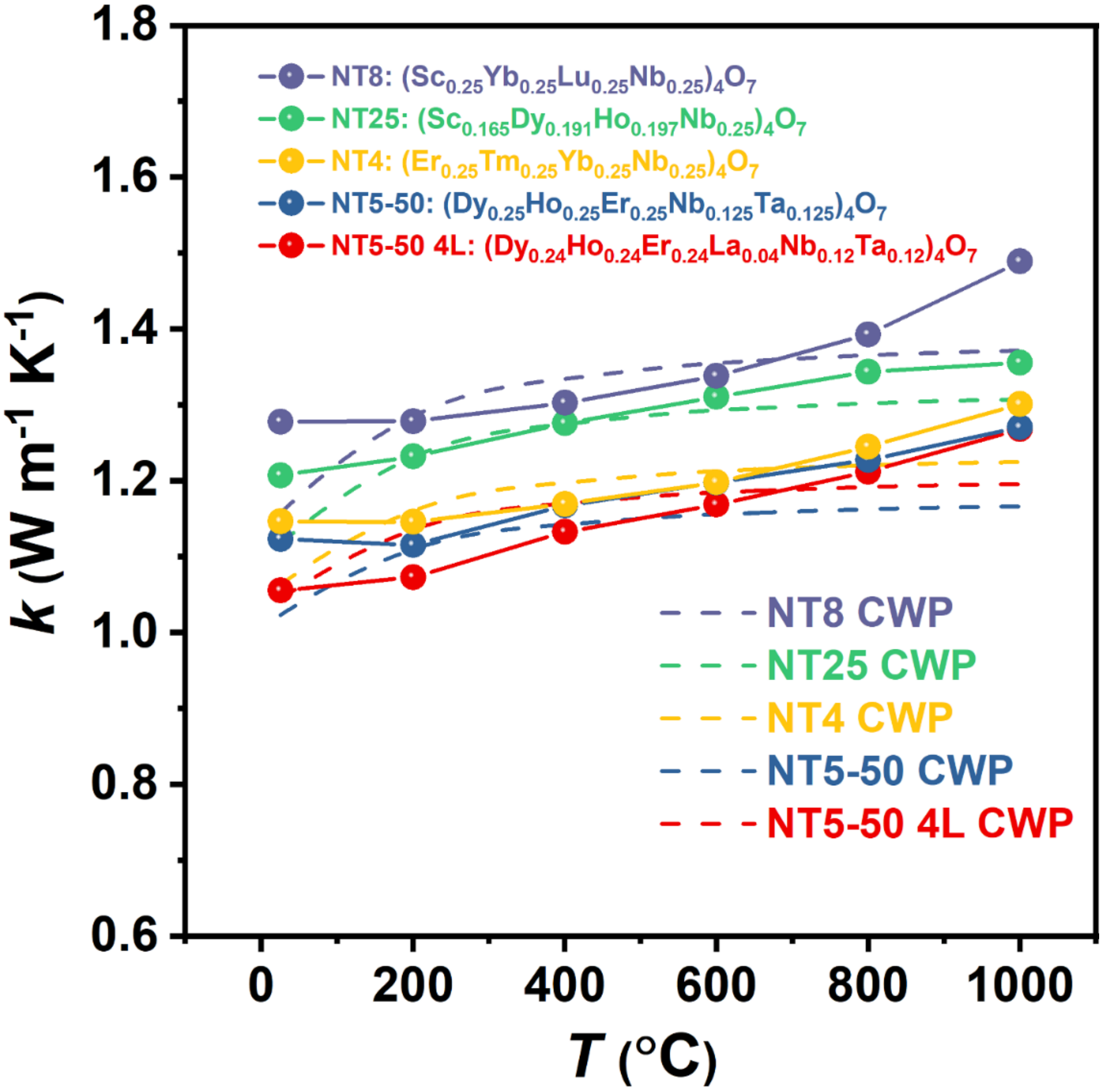

**Figure 4.** Temperature-dependent thermal conductivity of the five selected specimens (NT8, NT25, NT4, NT5-50, and NT5-50 4L) from room temperature to 1000°C. The $k_{min}$ as described by Cahill, Watson, and Pohl are shown as a dashed line [98].

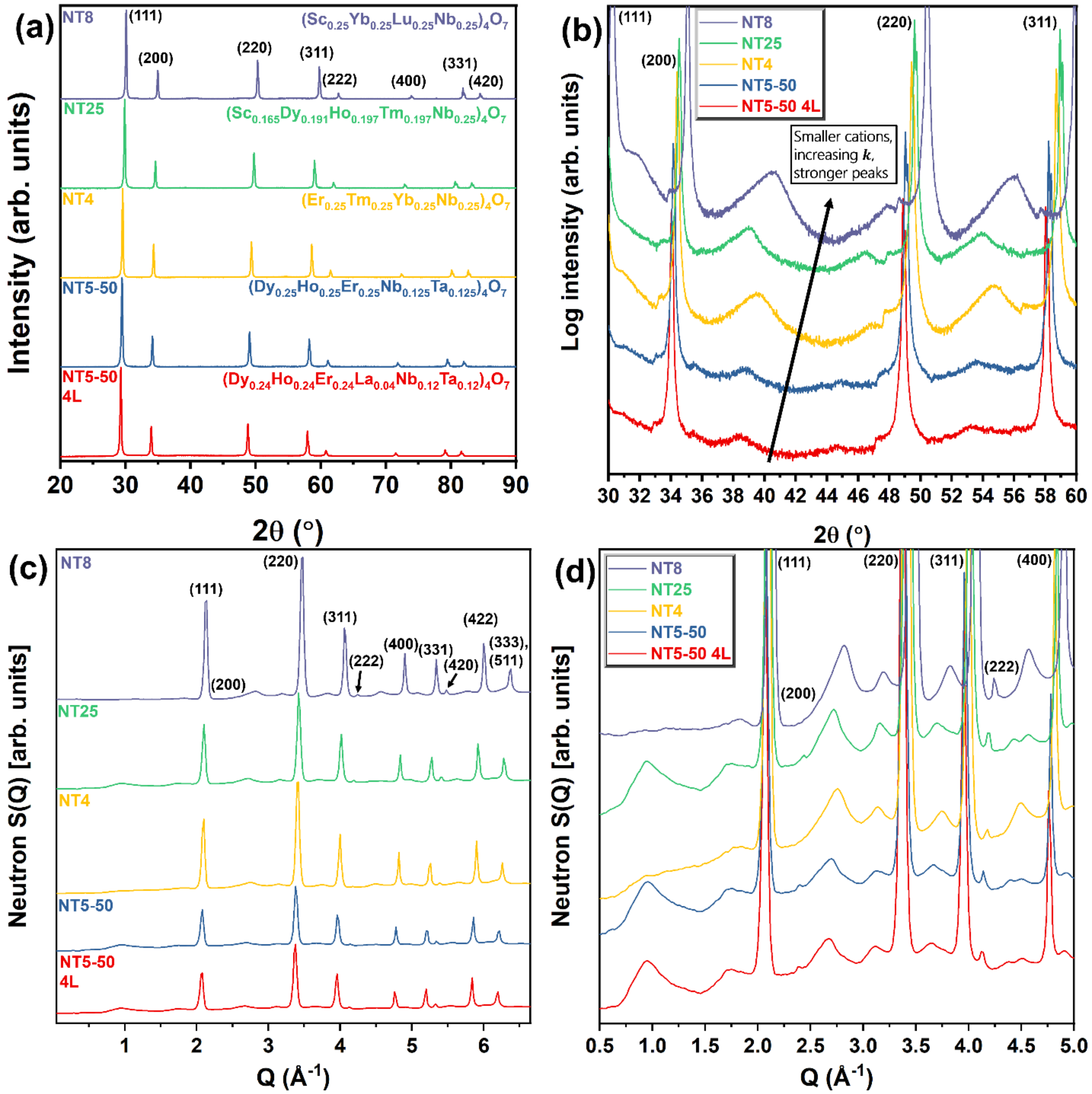

**Figure 5.** Benchtop X-ray diffraction patterns of the five selected specimens on a **(a)** linear intensity scale and **(b)** a close-up of diffuse scattering present on a logarithmic intensity scale. **(c)** Neutron diffraction total scattering patterns of these five specimens and **(d)** the diffuse scattering at a higher magnification.

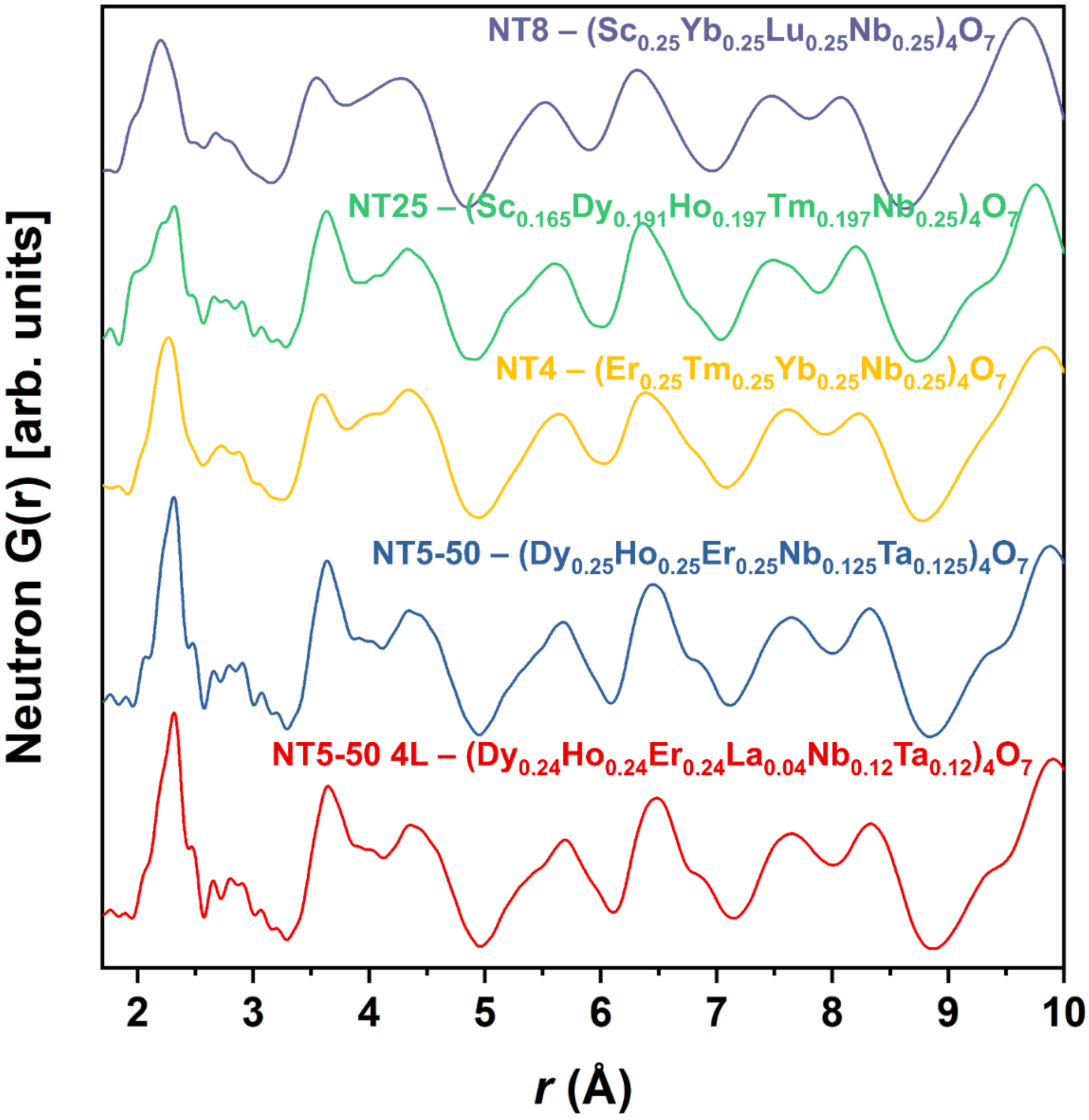

**Figure 6.** The partial distribution functions (PDFs, $G(r)$) of each of the five selected specimens.

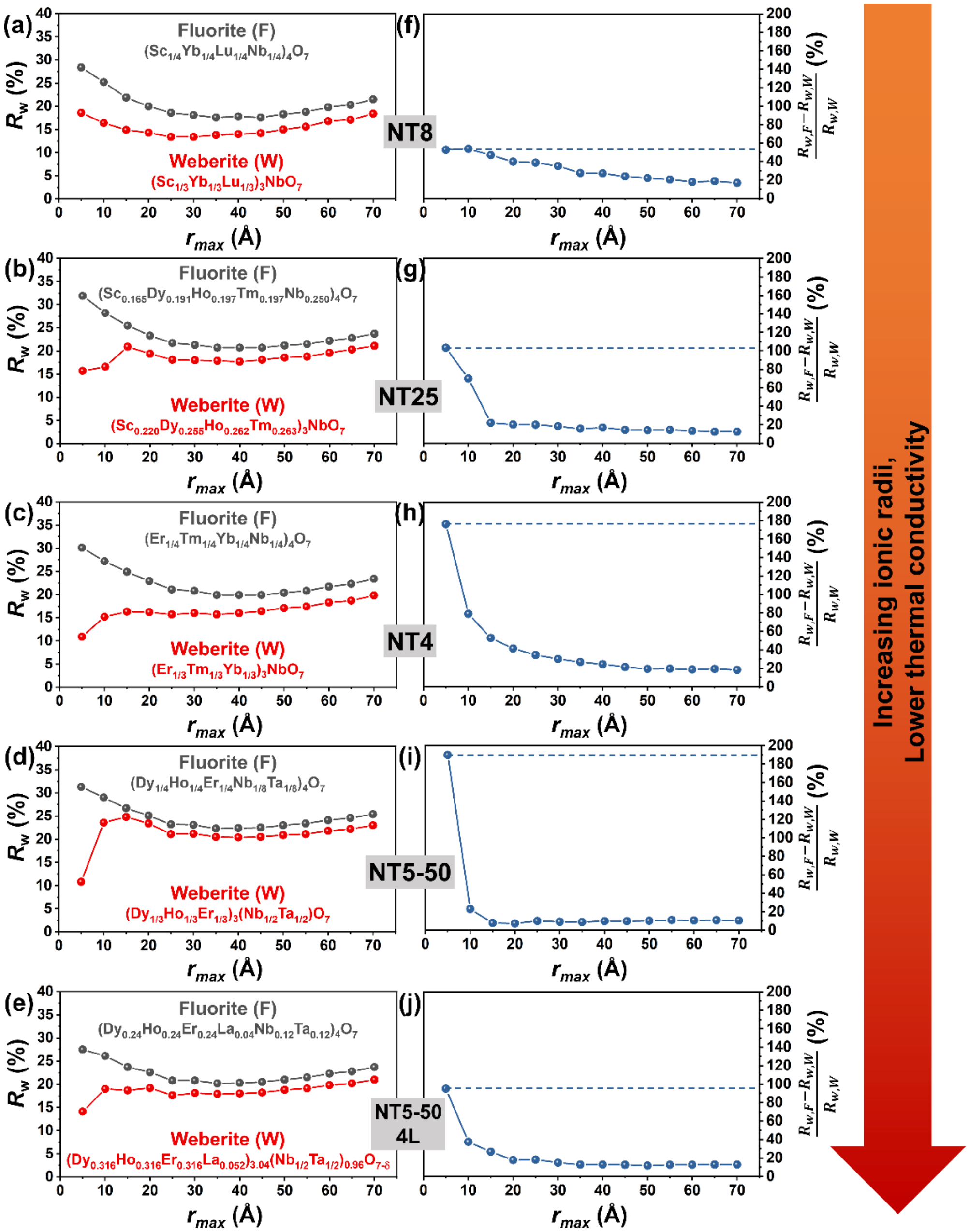

**Figure 7. (a-e)** The fitting parameter $R_w$ based on the upper bound of the fitting window for both the fluorite and weberite structures. The fitting began from $0.02 - 70$ Å and then was decreased in 5 Å steps to $0.02 - 5$ Å. After each step, the converged parameters from the previous step were used as the initial guess. **(f-j)** Percent difference between the fluorite and weberite $R_w$ factors. A stronger fit is found for the lower thermal conductivity specimens while also showing a steeper decline suggesting a smaller domain size for the weberite structure.

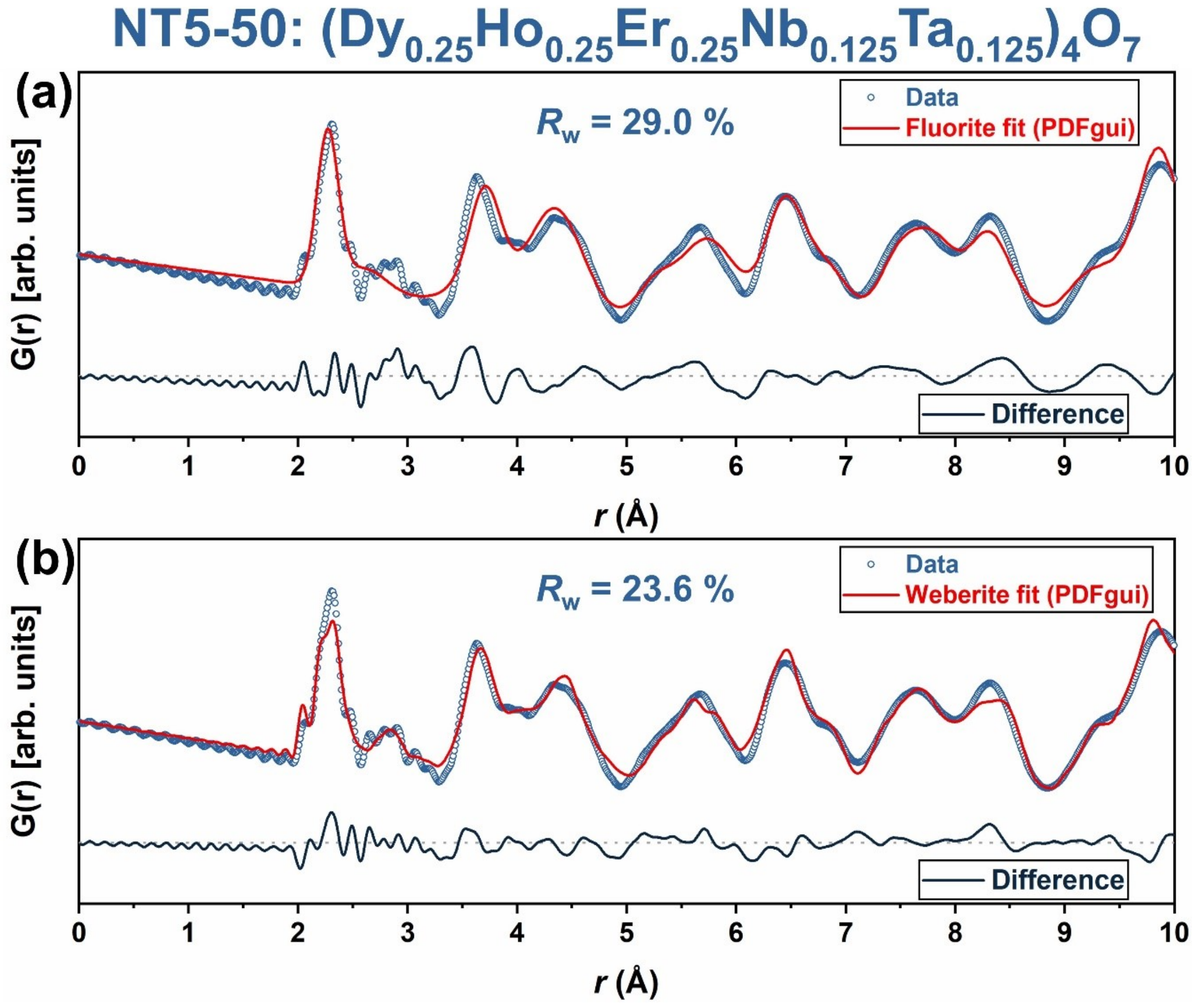

**Figure 8.** The fitting and difference curve of NT5-50 ($(Dy_{0.25}Er_{0.25}Ho_{0.25}Nb_{0.125}Ta_{0.125})_4O_7$) up to 10 Å using a **(a)** fluorite and **(b)** weberite structure using the PDFgui software. A better fit is achieved by assuming a weberite structure, particularly for the small peaks at $r < \sim 3.5$ Å.

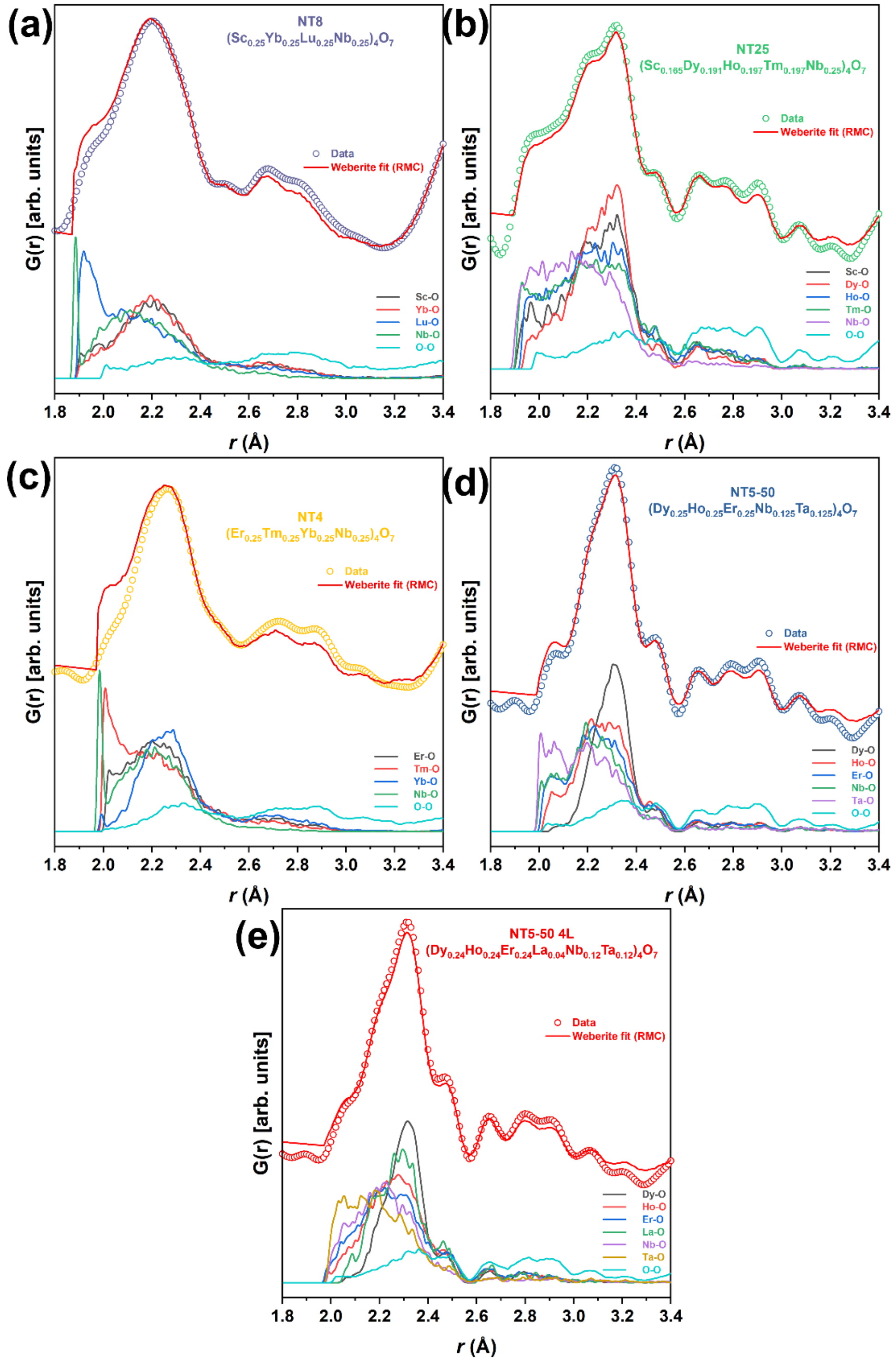

**Figure 9.** The fitting and the partial $G_{ij}(r)$ for each bonded pair and O-O of **(a)** NT8, **(b)** NT25, **(c)** NT4, **(d)** NT5-50, and **(e)** NT5-50 4L. All the specimens with Dy and Ho (NT25, NT5-50, and NT5-50) were found to have a stronger affinity for O ordering discernible by the sharper O-O interactions compared to NT8 and NT4.

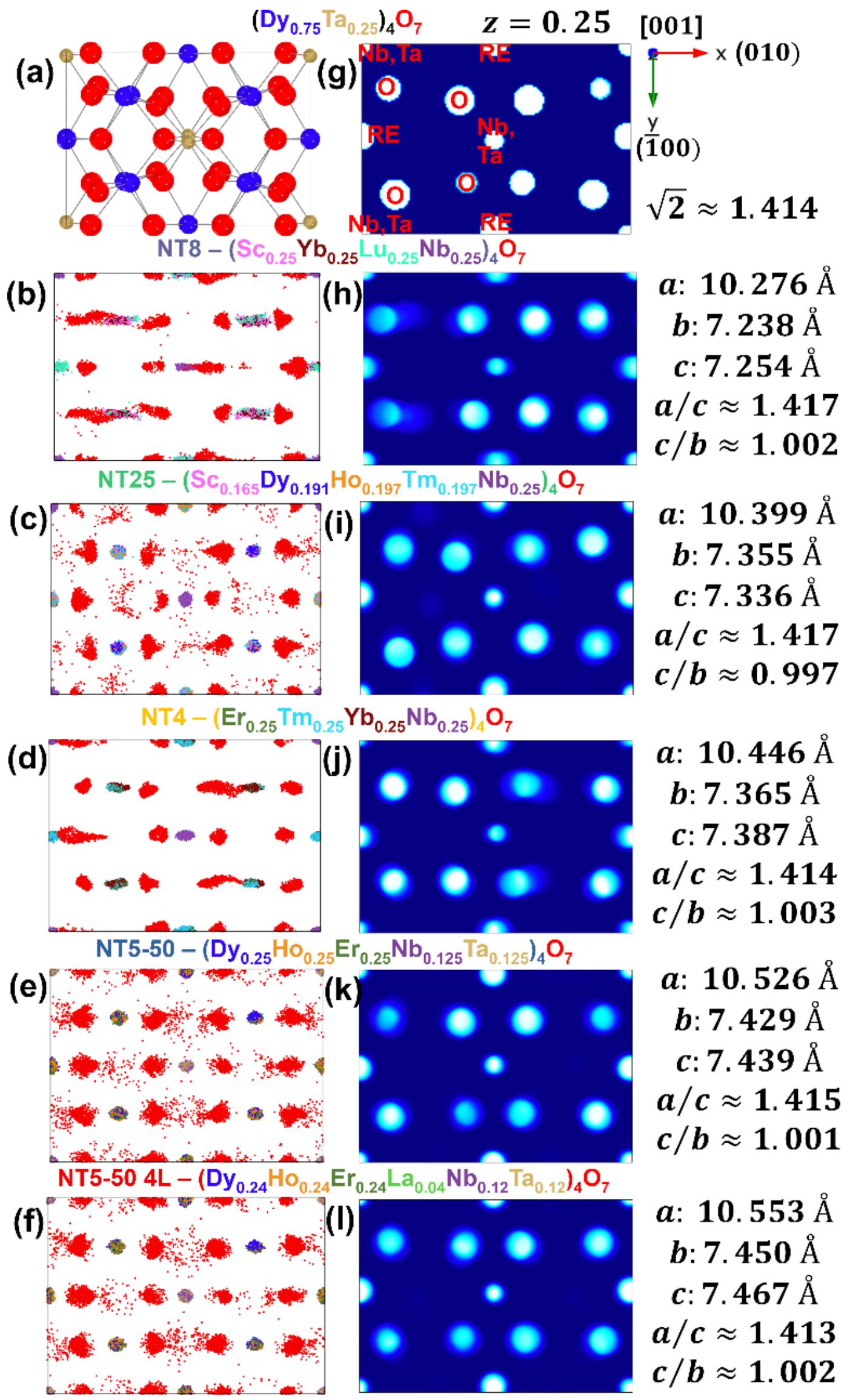

**Figure 10. (a)** Structural template ($Dy_3TaO_7$) of the weberite looking down the [001] plane. The collapsed unit cell of the converged reverse Monte Carlo simulation for **(b)** NT8, **(c)** NT25, **(d)** NT4, **(e)** NT5-50, and **(f)** NT5-50 4L in the [001] viewing direction. **(g)** Atomic density of $Dy_3TaO_7$ looking down the [001] plane at $z = 0.25$. **(h-l)** The atomic density heatmap of each specimen in the [001] viewing direction where $z = 0.25$ along with the converged lattice parameters. The compositions containing the larger Dy and Ho cations (NT25, NT5-50, and NT5-50 4L) have a more ordered O sublattice with some O atoms occupying interstitials while the compositions in absence of these larger cations shown stretched cation and anion positions in the (010) direction. The presence of large cations significantly affects the O sublattice but does not correlate well with the measured thermal conductivity. An alternative view of Figure 10 where the zone axis is slightly tilted in the +y direction to illustrate depth is provided in the Suppl. Fig. S17 in the SM.

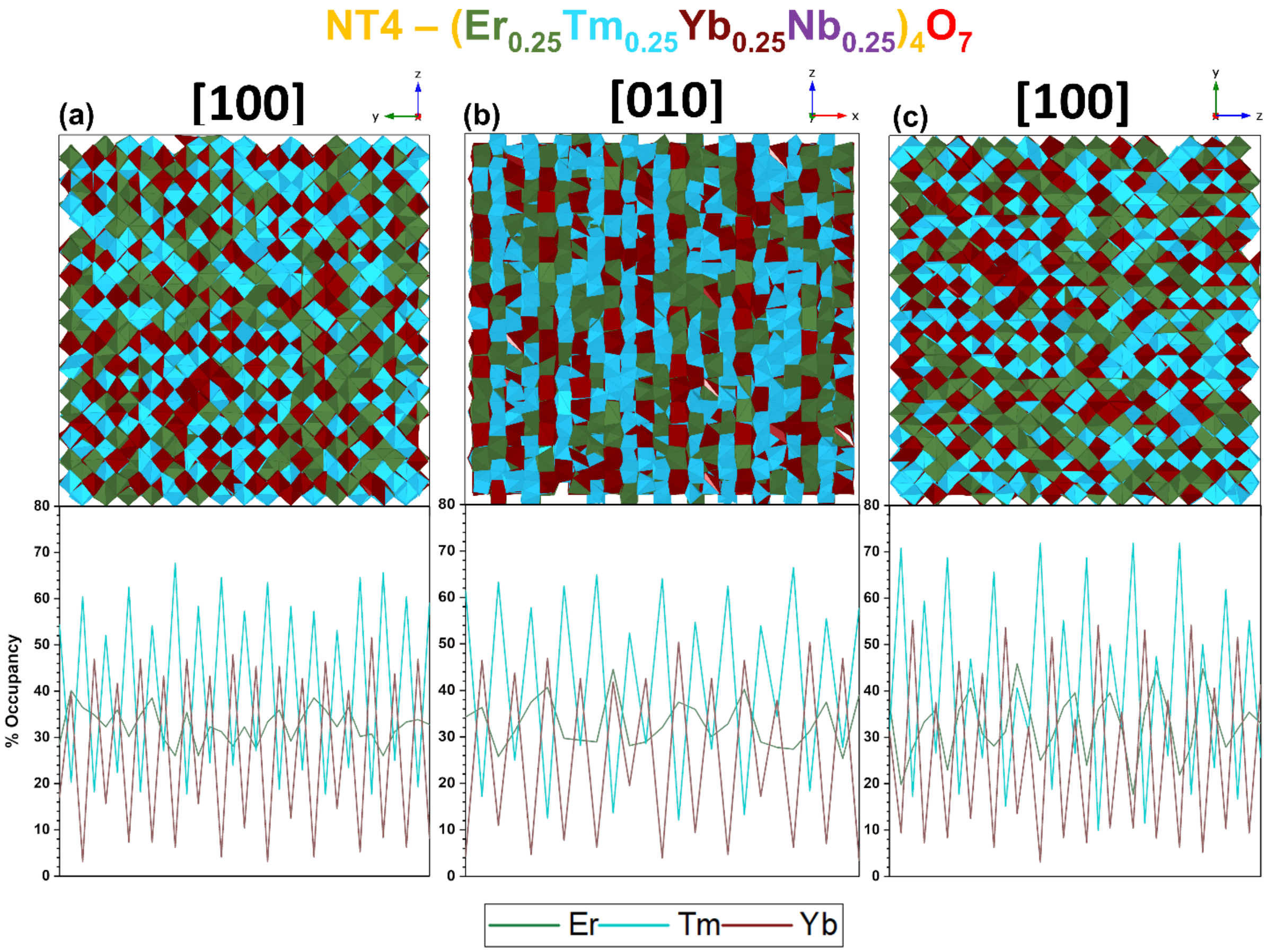

**Figure 11.** Polyhedral of the supercell and the elemental site distribution of Er, Tm, and Yb in NT4, $(Er_{0.25}Tm_{0.25}Yb_{0.25}Nb_{0.25})_4O_7$, along the **(a)** [100], **(b)** [010], and **(c)** [100] zone axes. See Suppl. Figs. S18-S21 in the SM for other four selected compositions.